\documentclass[preprint,preprintnumbers, prd, floatfix, superscriptaddress,nofootinbib] {revtex4-1}
\usepackage{epsfig}
\usepackage{subfigure}
\usepackage{dcolumn}
\usepackage{bm}
\usepackage[usenames ,dvipsnames]{xcolor}
\usepackage{slashed}
\usepackage{graphicx,color}
\begin{document}
\title{Study of $B^-\to \Lambda\bar p\eta^{(')}$ and 
$\bar B^0_s\to \Lambda\bar\Lambda\eta^{(')}$ decays}

\author{Y.K. Hsiao}
\affiliation{School of Physics and Information Engineering, Shanxi Normal University, Linfen 041004, China}

\author{C.Q. Geng}
\affiliation{School of Physics and Information Engineering, Shanxi Normal University, Linfen 041004, China}
\affiliation{Department of Physics, National Tsing Hua University, Hsinchu 300}
\affiliation{Chongqing University of Posts \& Telecommunications, Chongqing, 400065, China}

\author{Yu Yao}
\affiliation{Chongqing University of Posts \& Telecommunications, Chongqing, 400065, China}

\author{H.J. Zhao}
\affiliation{School of Physics and Information Engineering, Shanxi Normal University, Linfen 041004, China}

\date{\today}
\begin{abstract}
We study the three-body baryonic $B\to {\bf B\bar B'}M$ decays 
with $M$ representing the $\eta$ or $\eta'$ meson.
Particularly, we predict that ${\cal B}(B^-\to\Lambda\bar p\eta,\Lambda\bar p\eta')
=(5.3\pm 1.4,3.3\pm 0.7)\times 10^{-6}$ or $(4.0\pm 0.7,4.6\pm 1.1)\times 10^{-6}$, 
where the errors arise from the non-factorizable effects 
as well as the uncertainties 
in the $0\to {\bf B\bar B'}$ and $B\to{\bf B\bar B'}$ transition form factors,
while the two different results are due to overall relative signs between the form factors, 
causing  the constructive and destructive interference effects.
For the corresponding baryonic $\bar B_s^0$ decays, we find that 
${\cal B}(\bar B^0_s\to \Lambda\bar \Lambda \eta,\Lambda\bar \Lambda \eta')
=(1.2\pm 0.3,2.6\pm 0.8)\times 10^{-6}$ or $(2.1\pm 0.6,1.5\pm 0.4)\times 10^{-6}$
with the errors similar to those above.
The decays in question are accessible to the experiments at BELLE and LHCb.
\end{abstract}

\maketitle

\section{introduction}

In association with the QCD anomaly,
the $b$ and $c$-hadron decays with $\eta^{(\prime)}$ as the final states
have drawn lots of theoretical and experimental attentions, where
the $\eta$ and $\eta^\prime$ mesons are in fact 
the mixtures of the singlet $\eta_1$ and octet $\eta_8$ states, with
$\eta_{1,8}$ being decomposed as 
$\eta_n=(u\bar u+d\bar d)/\sqrt 2$ and $\eta_s=s\bar s$
in the FKS scheme~\cite{FKS}. 
In addition, the two configurations of
$b\to s n\bar n\to s\eta_n$  ($n=u$ or $d$)
and $b\to s\bar s s\to s\eta_s$  have been found to be the causes of
the dramatic interferences between 
the $B\to K^{(*)}\eta$ and $B\to K^{(*)}\eta^{\prime}$ decays,
that is,
${\cal B}(B\to K\eta)\ll {\cal B}(B\to K\eta')$ and 
${\cal B}(B\to K^*\eta)\gg {\cal B}(B\to K^*\eta')$~\cite{pdg}.
Note that the theoretical prediction gives 
${\cal B}(\bar B^0_s\to \eta^{(\prime)}\eta^\prime)\gg 
{\cal B}(\bar B^0_s\to \eta\eta)$~\cite{Hsiao:2015iiu},
while the only observation is 
${\cal B}(\bar B^0_s\to \eta^\prime\eta^\prime)=(3.3\pm 0.7)\times 10^{-5}$~\cite{Aaij:2015qga}.
On the other hand, 
with the dominant $b\to s\bar s s\to s\eta_s$ transition, 
the theoretical calculations result in
${\cal B}(\Lambda_b\to \Lambda\eta)\simeq 
{\cal B}(\Lambda_b\to \Lambda\eta')$~\cite{Ahmady:2003jz,Geng:2016gul},
which has not been confirmed by the the current data~\cite{LbtoLeta}.
For the dominant tree-level decay modes,
the theoretical results indicate that
${\cal B}(B\to \pi\eta)\simeq {\cal B}(B\to \pi\eta')$~\cite{Hsiao:2015iiu} and
${\cal B}(\Lambda_c^+\to p\eta)\simeq 
{\cal B}(\Lambda_c^+\to p\eta')$~\cite{Geng:2017esc,Geng:2018plk}.
Nonetheless, the observed values of ${\cal B}(B\to \pi\eta,\pi\eta')$ 
show a slight tension with the predictions. 
Experimentally, 
there are more to-be-measured decays with $\eta^{(\prime)}$, such as
the $B$ decays of $\bar B^0_s\to \eta\eta,\eta\eta'$ and
$\Lambda_c^+$ decays of $\Lambda_c^+\to p\eta^\prime,\Sigma^+\eta^\prime$.

Although
the charmless three-body baryonic $B$ decays ($B\to {\bf B\bar B'}M$) 
have been abundantly measured~\cite{pdg}, and well 
studied with the factorization~\cite{Hou:2000bz,Chua:2002wn,Chua:2002yd,
Geng:2005wt,Geng:2005fh,Geng:2006wz,Geng:2006jt,Chen:2008sw,
Hsiao:2016amt,Geng:2016fdw,Geng:2011pw,Hsiao:2017nga},
neither theoretical calculation nor experimental measurement 
for $B\to {\bf B\bar B'}\eta^{(\prime)}$ has been done yet.
We note that the prediction of 
${\cal B}(B^-\to\Lambda\bar p\phi)
=(1.5\pm 0.3)\times 10^{-6}$~\cite{Geng:2011pw} based on the factorization method
is  slightly larger than  the recent BELLE data of 
$(0.818\pm 0.215\pm 0.078)\times 10^{-6}$~\cite{Lu:2018qbw}.
In $B\to {\bf B\bar B'}M$,
the threshold enhancement has been observed 
as a generic feature~\cite{Abe:2002ds,Wang:2007as,
Aaij:2017vnw,Wei:2007fg,Bevan:2014iga,2b_baryonic},
which is shown as the peak at the threshold area of 
$m_{\bf B\bar B'}\simeq m_{\bf B}+m_{\bf B'}$ in the spectrum, 
with $m_{\bf B\bar B'}$ denoted as the invariant mass of the di-baryon.
With the threshold effect, one expects that 
${\cal B}(B\to{\bf B\bar B'}\eta^{(\prime)})\sim 10^{-6}$,
being accessible to the BELLE and LHCb experiments.
Furthermore, with  $b\to sn\bar n\to s\eta_n$ and $b\to s\bar s s\to s\eta_s$,
it is worth to explore if 
$B^-\to \Lambda\bar p\eta^{(')}$ and 
$\bar B^0_s\to \Lambda\bar\Lambda\eta^{(')}$ have
the interference effects for the branching ratios, which can be useful
to improve the knowledge of the underlying QCD anomaly
for the $\eta-\eta'$ mixing.
In this report, we will study the three-body baryonic $B$ decays
with one of the final states to be the $\eta$ or $\eta'$ meson state, where
the possible interference effects from the 
$b\to s n\bar n\to s\eta_n$ and $b\to s\bar s s\to s\eta_s$ transitions
can be investigated.

\section{Formalism}
\begin{figure}[t!]
\centering
\includegraphics[width=6in]{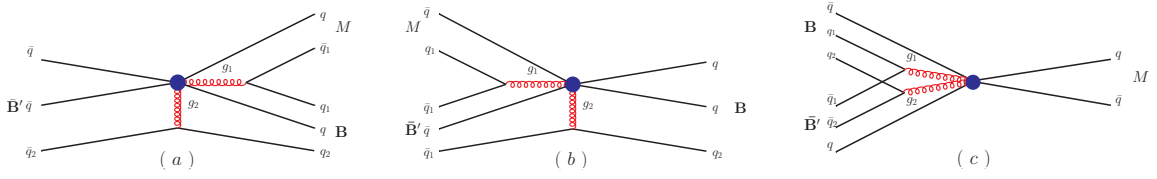}
\caption{The short-distance pictures for $B\to{\bf B\bar B'}M$ 
through the three quasi-two-body decays, where (a), (b) and (c) correspond to the collinearly moving 
$M{\bf B}$, $M{\bf \bar B'}$ and ${\bf B\bar B'}$,
respectively.}\label{SD_picture}
\end{figure}
%
Unlike the two-body mesonic $B\to MM$ decays, 
the $B\to {\bf B\bar B'}M$ decays require two additional quark pairs
for the $\bf B\bar B'$ formation.
This is in accordance with 
the short-distance pictures depicted in Fig.~\ref{SD_picture}~\cite{2b_baryonic,Suzuki:2006nn}, 
where $q_1\bar q_1$ and $q_2\bar q_2$
are connected by the gluons $g_{1,2}$, respectively.
In Fig.~\ref{SD_picture}a(b), the meson and (anti)baryon move collinearly, 
with $g_1$ for a collinear quark pair. 
By connecting to a back-to-back $q_2\bar q_2$ pair,
$g_2$ is far off the mass shell, such that it is a hard gluon,
resulting in the suppression with the factor of order $\alpha_s/q^2$.
There remain the resonant contributions observed to be small, 
which correspond to the suppression due to the short-distance pictures.
For example, one has 
${\cal B}(B^-\to p\Theta(1710)^{--},\Theta(1710)^{--}\to \bar p K^-)<9.1\times 10^{-8}$ and
${\cal B}(\bar B^0\to p\Theta(1540)^-,
\Theta(1540)^-\to \bar p K_s^0)<5\times 10^{-8}$~\cite{pdg}. 
Moreover,
$B^-\to \Lambda(1520)\bar p,\Lambda(1520)\to p K^-$ 
is observed with ${\cal B}\sim 10^{-7}$~\cite{Aaij:2013fla,Aaij:2014tua}.

On the other hand, 
the baryon pair in Fig.~\ref{SD_picture}c moves collinearly,
so that $g_{1,2}$ are both close to the mass shell, 
causing no suppression.  
Besides, the amplitudes can be factorized as
${\cal A}_1\propto\langle {\bf B\bar B'}|J_a|0\rangle \langle M|J_b|B\rangle$ and
${\cal A}_2\propto \langle M|J_a|0\rangle \langle {\bf B\bar B'}|J_b|B\rangle$.
%
\begin{figure}[t!]
\centering
\includegraphics[width=1.4in]{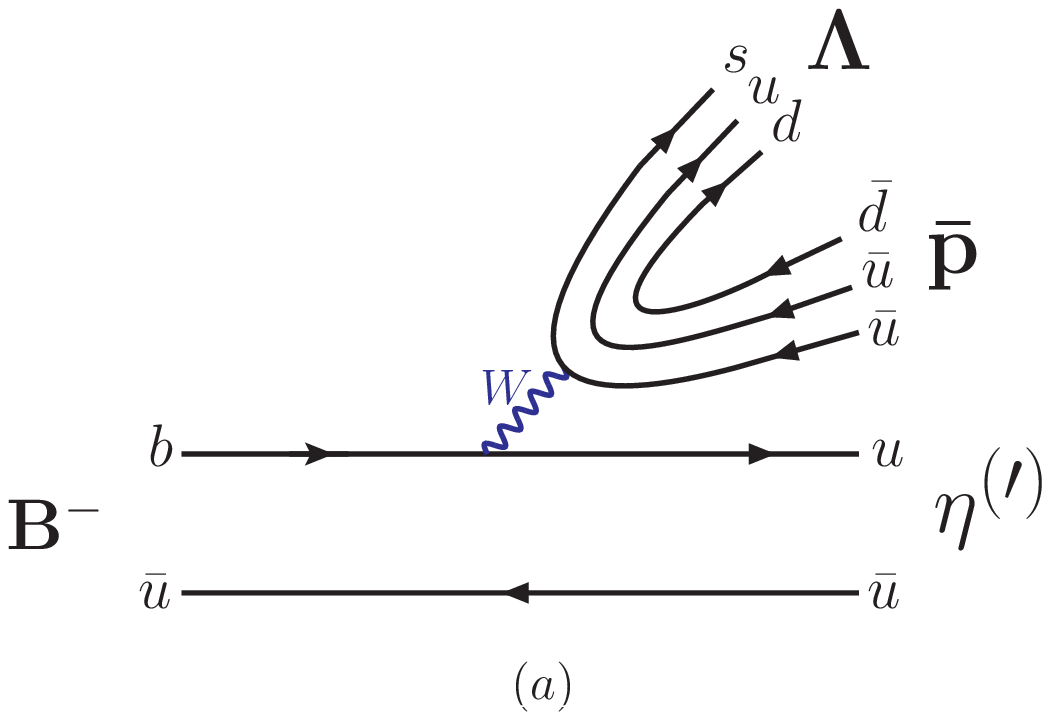}
\includegraphics[width=1.4in]{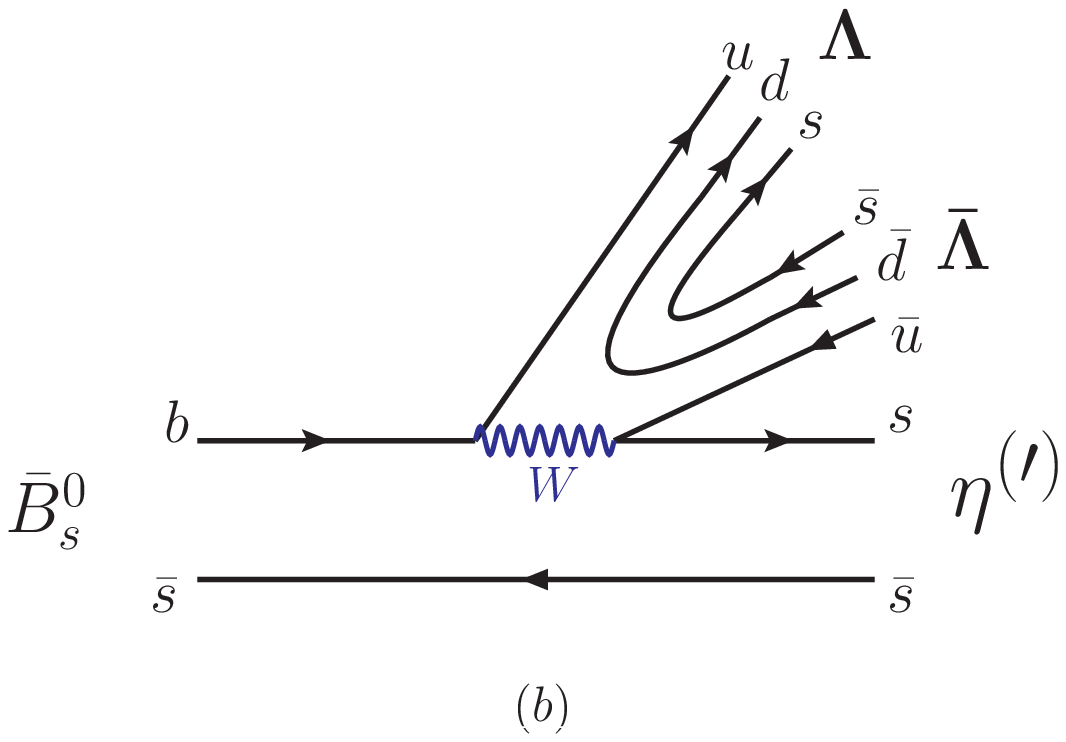}
\includegraphics[width=1.4in]{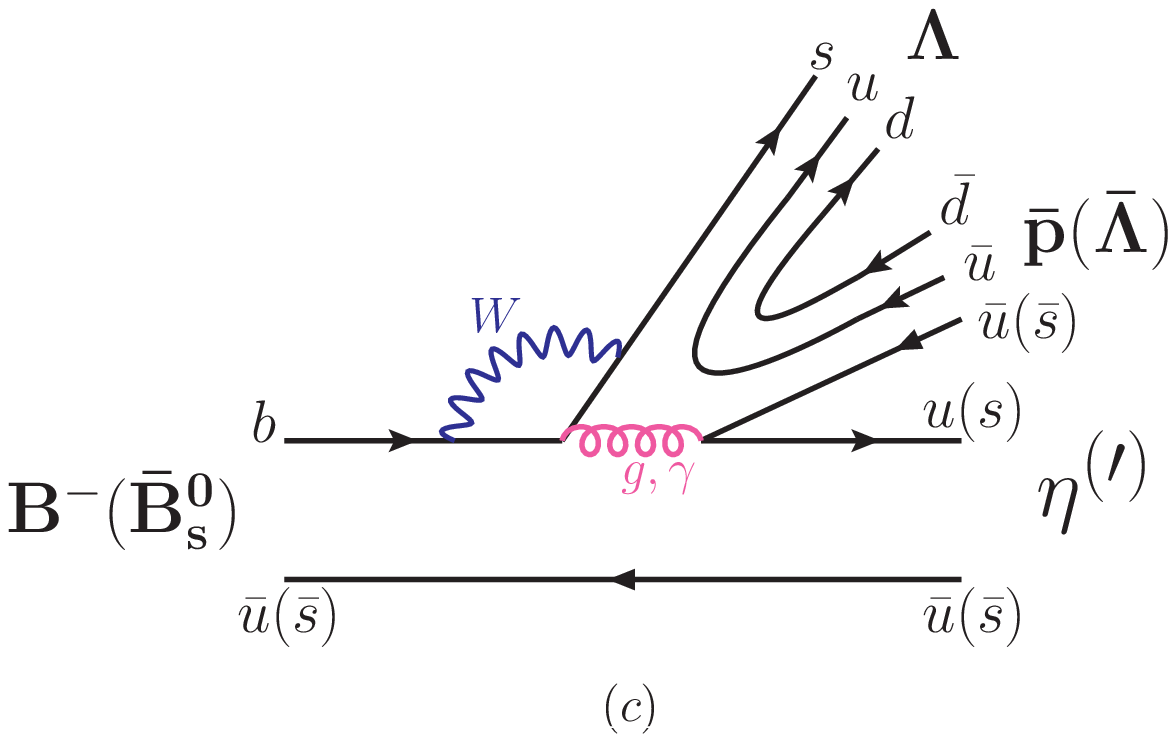}
\includegraphics[width=1.4in]{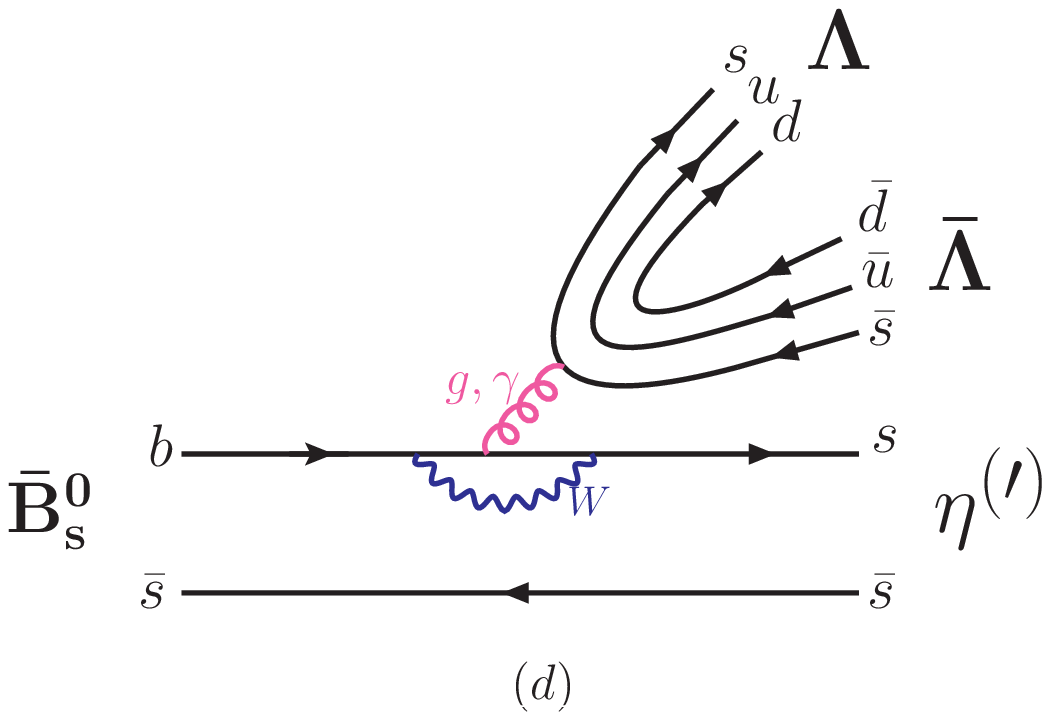}
\includegraphics[width=1.6in]{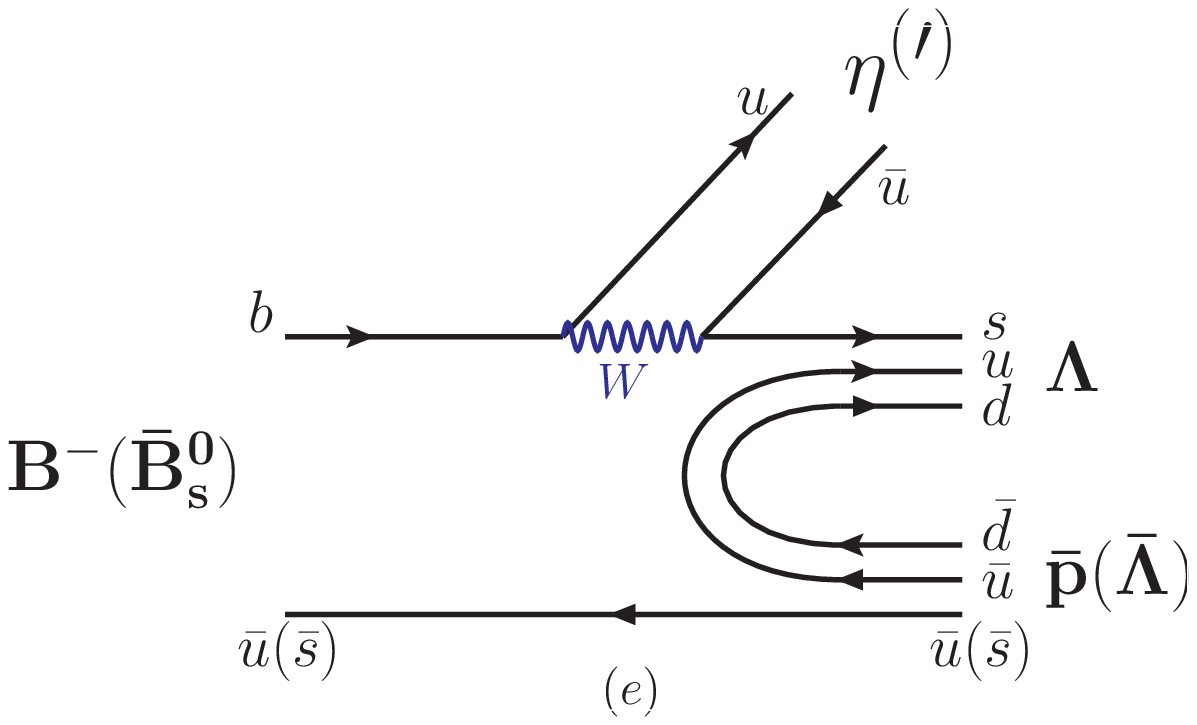}
\includegraphics[width=1.6in]{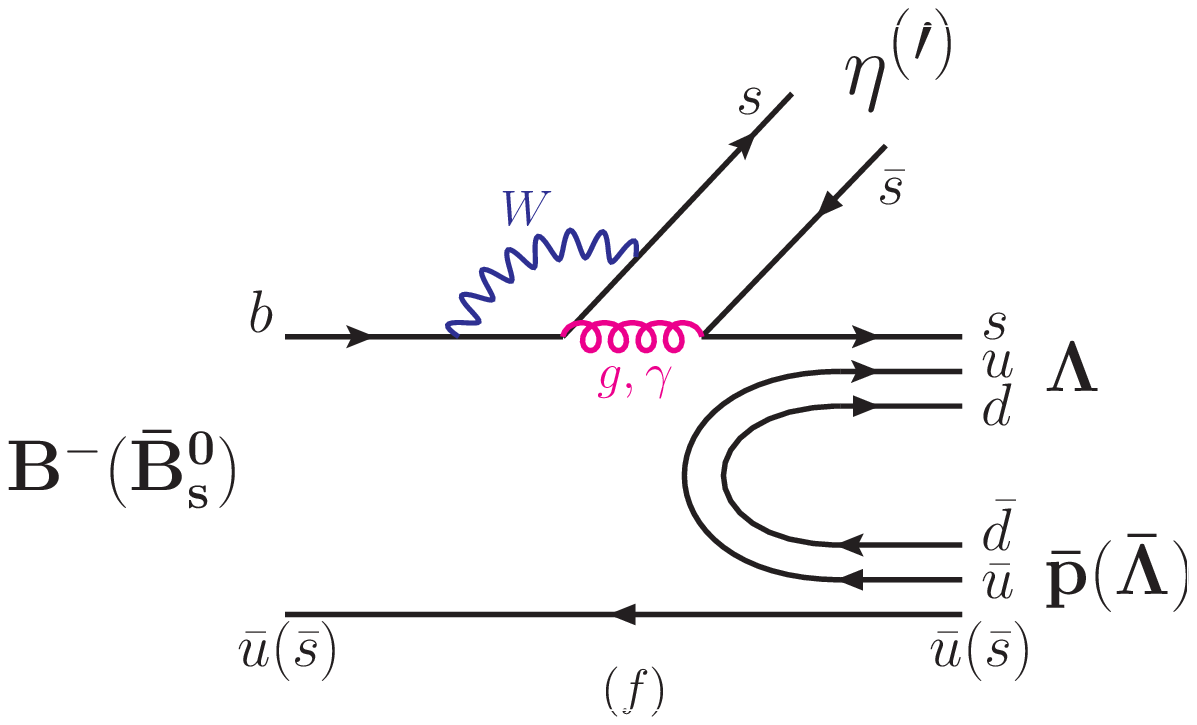}
\includegraphics[width=1.6in]{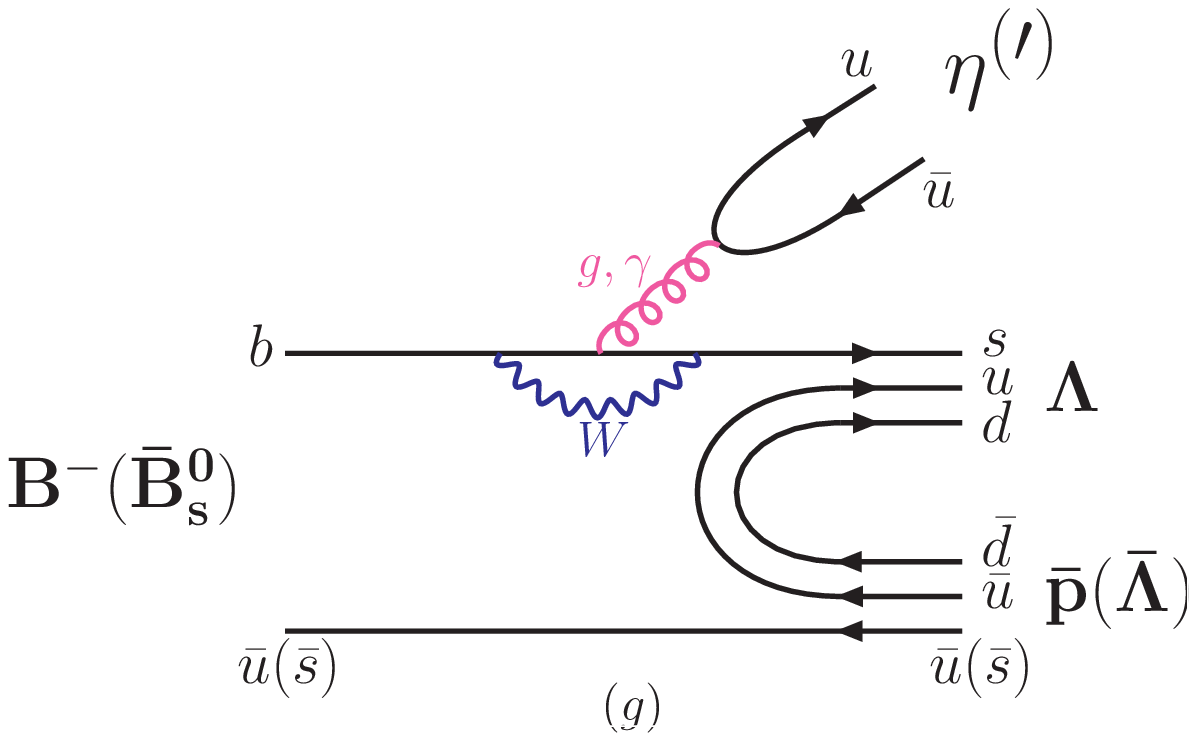}
\caption{Feynman diagrams for 
$B^-\to \Lambda\bar p\eta^{(\prime)}$ and 
$\bar B^0_s\to \Lambda\bar \Lambda\eta^{(\prime)}$decays
through (a,b,c,d)  $B\to \eta^{(\prime)}$ transitions with $0\to {\bf B\bar B'}$ productions
and
(e,f,g) $B\to {\bf B\bar B'}$ transitions with the recoiled $\eta^{(\prime)}$.}\label{dia}
\end{figure}
%
Accordingly, the  Feynman diagrams for
the three-body baryonic $B\to{\bf B\bar B'}\eta^{(\prime)}$ decays
with the short-distance approximation are shown in Fig.~\ref{dia}.
In our calculation, we use the generalized factorization as the theoretical approach.
The non-factorizable effects are included 
by the effective Wilson coefficients~\cite{Beneke,Cheng:2001tr,ali}.
In terms of the effective Hamiltonian 
for the $b\to sq\bar q$ transitions~\cite{Buras:1998raa},
the decay amplitudes of $B^-\to \Lambda\bar p\eta^{(\prime)}$
by the factorization can be derived as~\cite{ali,Geng:2006wz,Chua:2002wn,Chua:2002yd,
Geng:2005wt,Geng:2006jt,Hsiao:2016amt,Geng:2016fdw}
\begin{eqnarray}\label{amp1}
{\cal A}(B^-\to \Lambda\bar p\eta^{(\prime)})&=&
{\cal A}_1(B^-\to \Lambda\bar p\eta^{(\prime)})+{\cal A}_2(B^-\to \Lambda\bar p\eta^{(\prime)})\,,\nonumber\\
{\cal A}_1(B^-\to \Lambda\bar p\eta^{(\prime)})&=&\frac{G_F}{\sqrt 2}\bigg\{
\alpha_1\langle \Lambda\bar p|(\bar s\gamma_\mu(1-\gamma_5) u|0\rangle 
\langle \eta^{(\prime)}|\bar u\gamma_\mu(1-\gamma_5) b|B^-\rangle\nonumber\\
&+&\alpha_6\langle \Lambda\bar p|\bar s(1+\gamma_5) u|0\rangle
\langle \eta^{(\prime)}|\bar u(1-\gamma_5) b|B^-\rangle\bigg\}\,,\nonumber\\
{\cal A}_2(B^-\to \Lambda\bar p\eta^{(\prime)})&=&\frac{G_F}{\sqrt 2}\bigg\{ 
\bigg[\beta_2\langle \eta^{(\prime)}|\bar n\gamma_\mu\gamma_5 n|0\rangle+
\beta_3\langle \eta^{(\prime)}|\bar s \gamma_\mu \gamma_5 s|0\rangle\bigg]
\langle \Lambda\bar p|\bar s \gamma_\mu(1-\gamma_5) b|B^-\rangle\nonumber\\
&+&
\beta_6\langle \eta^{(\prime)}|\bar s\gamma_5 s|0\rangle
\langle \Lambda\bar p|\bar s(1-\gamma_5) b|B^-\rangle
\bigg\}\;,
\end{eqnarray}
where $n=u$ or $d$, $G_F$ is the Fermi constant, and  
${\cal A}_{1}$ and ${\cal A}_{2}$ correspond to
the two different decaying configurations in Fig.~\ref{dia}.
Similarly, the amplitudes of $\bar B^0_s\to \Lambda\bar \Lambda\eta^{(\prime)}$
are given by
\begin{eqnarray}\label{amp1b}
{\cal A}(\bar B^0_s\to \Lambda\bar \Lambda\eta^{(\prime)})&=&
{\cal A}_1(\bar B^0_s\to \Lambda\bar \Lambda\eta^{(\prime)})+
{\cal A}_2(\bar B^0_s\to \Lambda\bar \Lambda\eta^{(\prime)})\,,\nonumber\\
{\cal A}_1(\bar B^0_s\to \Lambda\bar \Lambda\eta^{(\prime)})&=&
\frac{G_F}{\sqrt 2}\bigg\{ 
\bigg[
\langle  \Lambda\bar \Lambda|\bar n\gamma_\mu(\alpha_2^+ -\alpha_2^-\gamma_5) n|0\rangle+
\langle  \Lambda\bar \Lambda|\bar s\gamma_\mu(\alpha_3^+ -\alpha_3^-\gamma_5) s
|0\rangle\bigg]\nonumber\\
&\times& \langle \eta^{(\prime)}|\bar s \gamma_\mu(1-\gamma_5) b|\bar B^0_s\rangle
+\alpha_6^s\langle \Lambda\bar \Lambda|\bar s(1+\gamma_5) s|0\rangle
\langle \eta^{(\prime)}|\bar s(1-\gamma_5) b|\bar B^0_s\rangle\bigg\}\;,\nonumber\\
%
{\cal A}_2(\bar B^0_s\to \Lambda\bar \Lambda\eta^{(\prime)})&=&\frac{G_F}{\sqrt 2}\bigg\{ 
\bigg[\beta_2\langle \eta^{(\prime)}|\bar n\gamma_\mu\gamma_5 n|0\rangle+
\beta_3\langle \eta^{(\prime)}|\bar s \gamma_\mu \gamma_5 s|0\rangle\bigg]
\langle \Lambda\bar \Lambda|\bar s \gamma_\mu(1-\gamma_5) b|\bar B^0_s\rangle\nonumber\\
&+&
\beta_6\langle \eta^{(\prime)}|\bar s\gamma_5 s|0\rangle
\langle \Lambda\bar \Lambda|\bar s(1-\gamma_5) b|\bar B^0_s\rangle
\bigg\}\;.
\end{eqnarray}
The parameters $\alpha_i$ and $\beta_i$ in Eqs.~(\ref{amp1}) and (\ref{amp1b})
are defined as
\begin{eqnarray}\label{alpha_beta}
\alpha_1&=&V_{ub}V_{us}^* a_1-V_{tb}V_{ts}^*(a_4+a_{10})\,,\nonumber\\
\alpha_2^\pm&=&
V_{ub}V_{us}^* a_2-V_{tb}V_{ts}^* (2a_3\pm 2a_5\pm\frac{a_7}{2}+\frac{a_9}{2})\,,\nonumber\\
\alpha_3^\pm&=&
-V_{tb}V_{ts}^*(a_3+a_4\pm a_5\mp\frac{a_7}{2}-\frac{a_9}{2}-\frac{a_{10}}{2})\,,\nonumber\\
\alpha_6&=&V_{tb}V_{ts}^*2(a_6+a_8)\,,\nonumber\\
\alpha_6^s&=&V_{tb}V_{ts}^*2(a_6-\frac{a_8}{2})\,,\nonumber\\
\beta_2&=&-\alpha_2^-\;,\beta_3=-\alpha_3^-\;,\beta_6=\alpha_6^s\;, 
\end{eqnarray}
where $V_{ij}$ the CKM matrix elements, and
$a_i=c^{eff}_i+c^{eff}_{i\pm 1}/N_c$ for $i=$odd (even) 
with $N_c$ the effective color number in the generalized factorization approach, 
consisting of the effective Wilson coefficients $c_i^{eff}$~\cite{ali}.
The matrix elements in Eq.~(\ref{amp1}) 
for the $\eta^{(\prime)}$ productions read~\cite{Beneke:2002jn}
\begin{eqnarray}
&&\langle \eta^{(\prime)}|\bar n\gamma_\mu \gamma_5 n|0\rangle=
-\frac{i}{\sqrt 2}f^n_{\eta^{(\prime)}} q_\mu\,,\nonumber\\
&&\langle \eta^{(\prime)}|\bar s\gamma_\mu \gamma_5 s|0\rangle=-if^s_{\eta^{(\prime)}} q_\mu\,,\nonumber\\
&&2m_s \langle \eta^{(\prime)}|\bar s\gamma_5 s|0\rangle=-ih^s_{\eta^{(\prime)}}\,,
\end{eqnarray}
with $f^{n,s}_{\eta^{(\prime)}}$ and $h^s_{\eta^{(\prime)}}$ the decay constants
and $q_\mu$ the four-momentum vector. The $\eta$ and $\eta'$ meson states 
mix with $|\eta_n\rangle=(|u\bar u+d\bar d\rangle)/\sqrt 2$ 
and $|\eta_s\rangle=|s\bar s\rangle$~\cite{FKS},
in terms of the mixing matrix:
\begin{eqnarray}\label{mixing}
\left(\begin{array}{c}
\eta\\
\eta'
\end{array}\right)
=
\left(\begin{array}{cc}
\cos\phi&-\sin\phi\\
\sin\phi&\cos\phi
\end{array}\right)
\left(\begin{array}{c}
\eta_n\\
\eta_s
\end{array}\right)\;,
\end{eqnarray}
with the mixing angle $\phi=(39.3\pm 1.0)^\circ$.
Therefore, 
$f^n_{\eta^{(\prime)}}$ and $f^s_{\eta^{(\prime)}}$ actually 
come from $f_n$ and $f_s$ for $\eta_n$ and $\eta_s$, respectively. In addition,
$h^s_{\eta^{(\prime)}}$ receive the contributions from the QCD anomaly~\cite{Beneke:2002jn}.
The matrix elements of 
the $B\to \eta^{(\prime)}$ transitions are parameterized as~\cite{BSW}
\begin{eqnarray}\label{ff1}
\langle \eta^{(\prime)}| \bar q \gamma^\mu b|B\rangle&=&
\bigg[(p_B+p_{\eta^{(\prime)}})^\mu-\frac{m^2_B-m^2_{\eta^{(\prime)}}}{t}q^\mu\bigg]
F_1^{B\eta^{(\prime)}}(t)+\frac{m^2_B-m^2_{\eta^{(\prime)}}}{t}q^\mu F_0^{B\eta^{(\prime)}}(t)\,,
\end{eqnarray}
with $q=p_B-p_{\eta^{(\prime)}}=p_{\bf B}+p_{\bf\bar B'}$ and $t\equiv q^2$,
where the momentum dependences 
are expressed as~\cite{MFD}
\begin{eqnarray}\label{form2}
F^{B\eta^{(\prime)}}_1(t)=
\frac{F^{B\eta^{(\prime)}}_1(0)}{(1-\frac{t}{M_V^2})
(1-\frac{\sigma_{11} t}{M_V^2}+\frac{\sigma_{12} t^2}{M_V^4})}\,,\;
F^{B\eta^{(\prime)}}_0(t)&=&\frac{F^{B\eta^{(\prime)}}_0(0)}
{1-\frac{\sigma_{01} t}{M_V^2}+\frac{\sigma_{02} t^2}{M_V^4}}\,.
\end{eqnarray}
According to the mixing matrix in Eq.~(\ref{mixing}),
one has 
\begin{eqnarray}\label{form2b}
(F^{B\eta},F^{B\eta'})&=&(F^{B\eta_n}\cos\phi,F^{B\eta_n}\sin\phi)\,,\nonumber\\
(F^{B_s\eta},F^{B_s\eta'})&=&(-F^{B_s\eta_s}\sin\phi,F^{B_s\eta_s}\cos\phi)\,,
\end{eqnarray}
for the $B^-$ and $\bar B^0_s$ transitions to $\eta^{(\prime)}$, respectively,
where $F^{B\eta^{(\prime)}}$ represent $F^{B\eta^{(\prime)}}_{1,0}(0)$.

The matrix elements in Eq.~(\ref{amp1}) for the baryon-pair productions
are parameterized as~\cite{Chua:2002yd,Geng:2005wt} 
\begin{eqnarray}\label{FFactor1}
\langle {\bf B\bar B'}|(\bar qq')_V|0\rangle
&=&
\bar u\bigg[F_1\gamma_\mu+\frac{F_2}{m_{\bf B}+m_{\bf \bar B'}}i\sigma_{\mu\nu}q^\nu\bigg]v\;,\nonumber\\
\langle {\bf B\bar B'}|(\bar qq')_A|0\rangle
&=&\bar u\bigg[g_A\gamma_\mu+\frac{h_A}{m_{\bf B}+m_{\bf \bar B'}}q_\mu\bigg]\gamma_5 v\,,\nonumber\\
\langle {\bf B\bar B'}|(\bar qq')_S|0\rangle &=&f_S\bar uv\;,\nonumber\\
\langle {\bf B\bar B'}|(\bar qq')_P|0\rangle&=&g_P\bar u \gamma_5 v\,,
\end{eqnarray}
with $(\bar qq')_{V,A,S,P}=
(\bar q\gamma_\mu q',\bar q\gamma_\mu\gamma_5q',\bar qq',\bar q\gamma_5 q')$,
where $u$($v$) is the (anti-)baryon spinor, and  
$(F_{1,2},g_A,h_A,f_S,g_P)$ are the timelike baryonic form factors.
Meanwhile, the matrix elements of the $B\to{\bf B\bar B'}$ transitions
are written to be~\cite{Chua:2002wn,Geng:2006wz}
\begin{eqnarray}\label{FFactor2}
\langle {\bf B\bar B'}|(\bar sb)_V|B\rangle&=&
i\bar u[  g_1\gamma_{\mu}+g_2i\sigma_{\mu\nu}p^\nu +g_3 p_{\mu} 
+g_4q_\mu +g_5(p_{\bf\bar B'}-p_{\bf B})_\mu]\gamma_5v\,,\nonumber\\
\langle {\bf B\bar B'}|(\bar sb)_A|B\rangle&=&
i\bar u[ f_1\gamma_{\mu}+f_2i\sigma_{\mu\nu}p^\nu +f_3 p_{\mu} 
+f_4q_\mu +f_5(p_{\bf\bar B'}-p_{\bf B})_\mu]v\,,\nonumber\\
\langle {\bf B\bar B'}|(\bar s b)_S|B\rangle&=&
i\bar u[ \bar g_1\slashed p+\bar g_2(E_{\bf \bar B'}+E_{\bf B})
+\bar g_3(E_{\bf \bar B'}-E_{\bf B})]\gamma_5v\,,\nonumber\\
\langle {\bf B\bar B'}|(\bar sb)_P|B\rangle&=&
i\bar u[ \bar f_1\slashed p+\bar f_2(E_{\bf \bar B'}+E_{\bf B})
+\bar f_3(E_{\bf \bar B'}-E_{\bf B})]v\,,
\end{eqnarray}
with $p_\mu=(p_B-q)_\mu$,
where $g_i(f_i)$ $(i=1,2, ...,5)$ and $\bar g_j(\bar f_j)$ $(j=1,2,3)$
are the $B\to{\bf B\bar B'}$ transition form factors.
The momentum dependences of 
the baryonic form factors in Eqs.~(\ref{FFactor1}) and (\ref{FFactor2})
depend on the approach of perturbative QCD counting rules, 
given by~\cite{Brodsky:1973kr,Brodsky:2003gs,
Chua:2002wn,Geng:2006wz},
\begin{eqnarray}\label{timelikeF2}
&&F_1=\frac{\bar C_{F_1}}{t^2}\,,\;g_A=\frac{\bar C_{g_A}}{t^2}\,,\;
f_S=\frac{\bar C_{f_S}}{t^2}\,,\;g_P=\frac{\bar C_{g_P}}{t^2}\,,\;\nonumber\\
&&f_i=\frac{D_{f_i}}{t^3}\,,\;g_i=\frac{D_{g_i}}{t^3}\,,\;
\bar f_i=\frac{D_{\bar f_i}}{t^3}\,,\;\bar g_i=\frac{D_{\bar g_i}}{t^3}\,,
\end{eqnarray}
where $\bar C_i=C_i [\text{ln}({t}/{\Lambda_0^2})]^{-\gamma}$
with $\gamma=2.148$ and $\Lambda_0=0.3$ GeV. 
Compared to the $0\to{\bf B\bar B'}$ form factors, 
the $B\to{\bf B\bar B'}$  ones have an additional $1/t$,
which is for a gluon to speed up the slow spectator quark in $B$.
Due to
$F_2=F_1/(t\text{ln}[t/\Lambda_0^2])$ in~\cite{Belitsky:2002kj},
derived to be much less than $F_1$, and $h_A=C_{h_A}/t^2$~\cite{Hsiao:2014zza} 
that corresponds to the smallness of 
${\cal B}(\bar B^0\to p\bar p)\sim 10^{-8}$~\cite{Aaij:2013fta,Aaij:2017gum},
we neglect $F_2$ and $h_A$.
Under the $SU(3)$ flavor and $SU(2)$ spin symmetries,
the constants $C_i$ can be related, 
given by~\cite{Brodsky:1973kr,Chua:2002yd,Hsiao:2017nga}
\begin{eqnarray}\label{C||}
&&
(C_{F_1},C_{g_A},C_{f_S},C_{g_P})=
\sqrt\frac{3}{2}(C_{||},C_{||}^*,-\bar C_{||},-\bar C_{||}^*)\,,\;\;\;
\text{(for $\langle \Lambda\bar p|(\bar s u)_{V,A,S,P}|0\rangle$)}\nonumber\\
&&
(C_{F_1},C_{g_A},C_{f_S},C_{g_P})=
(C_{||},C_{||}^*,-\bar C_{||},-\bar C_{||}^*)\,,\;\;\;
\text{(for $\langle \Lambda\bar \Lambda|(\bar s s)_{V,A,S,P}|0\rangle$)}\nonumber\\
&&
(C_{F_1},C_{g_A})=\frac{1}{2}(C_{||}+C_{\overline{||}},C_{||}^*-C_{\overline{||}}^*)\,,\;\;\;
\text{(for $\langle \Lambda\bar \Lambda|(\bar n n)_{V,A}|0\rangle$)}
\end{eqnarray}
with $C_{||(\overline{||})}^*\equiv C_{||(\overline{||})}+\delta C_{||(\overline{||})}$
and $\bar C_{||}^*\equiv \bar C_{||}+\delta \bar C_{||}$, where
$\delta C_{||(\overline{||})}$ and $\delta \bar C_{||}$ are added 
to account for the broken symmetries,
indicated by the large and unexpected 
angular distributions in $\bar B^0\to \Lambda\bar p\pi^+$ and 
$B^-\to \Lambda\bar p\pi^0$~\cite{Wang:2007as}.
With the same symmetries~\cite{Chua:2002wn,Geng:2006wz,Geng:2006jt,
Hsiao:2016amt,Geng:2016fdw}, 
$D_i$ are related by\begin{eqnarray}\label{fitD1}
&&\text{$\langle \Lambda\bar p|(\bar s b)_{V,A}|B^-\rangle$:}\;\;
D_{g_1}=D_{f_1}=\sqrt\frac{3}{2}D_{||}\,,\;D_{g_{4,5}}=-D_{f_{4,5}}=-\sqrt\frac{3}{2}D_{||}^{4,5}\,,
\nonumber\\
&&\text{$\langle \Lambda\bar p|(\bar s b)_{S,P}|B^-\rangle$:}\;\;
D_{\bar g_1}=-D_{\bar f_1}=\sqrt\frac{3}{2}\bar D_{||}\,,\;
D_{\bar g_{2,3}}=D_{\bar f_{2,3}}=-\sqrt\frac{3}{2}\bar D_{||}^{2,3}\,,
\nonumber\\
&&\text{$\langle \Lambda\bar \Lambda|(\bar s b)_{V,A}|\bar B^0_s\rangle$:}\;\;
D_{g_1}=D_{f_1}=D_{||}\,,\;D_{g_{4,5}}=-D_{f_{4,5}}=-D_{||}^{4,5}\,,
\nonumber\\
&&\text{$\langle \Lambda\bar \Lambda|(\bar s b)_{S,P}|\bar B^0_s\rangle$:}\;\;
D_{\bar g_1}=-D_{\bar f_1}=\bar D_{||}\,,\;
D_{\bar g_{2,3}}=D_{\bar f_{2,3}}=-\bar D_{||}^{2,3}\,,
\end{eqnarray}
where the ignorances  of $D_{g_{2,3}}$ and $D_{f_{2,3}}$ 
correspond to the derivations of
$f_M p^\mu\bar u (\sigma_{\mu\nu} p^\nu)v=0$ for $g_2(f_2)$ and 
$f_M p^\mu \bar u p_\mu v\propto m_M^2$ for $f_3(g_3)$ in the amplitudes.
For the integration over the phase space in the three-body decay,
we refer the general equation of the decay width in the PDG, given by~\cite{pdg} 
\begin{eqnarray}\label{gamma1}
\Gamma=\int_{m_{12}^2}\int_{m_{23}^2}
\frac{1}{(2\pi)^3}\frac{|\bar {\cal A}|^2}{32M^3_B}dm_{12}^2 dm_{23}^2\,,
\end{eqnarray}
with $m_{12}=p_{\bf B}+p_{\bf\bar B'}$ and $m_{23}=p_{\bf B}+p_{\eta^{(\prime)}}$,
where $|\bar {\cal A}|^2$ represents the amplitude squared
with the total summations of the baryon spins. 
On the other hand, we can also  study the partial decay rate in terms of the 
 the angular  dependence, given by~\cite{Geng:2006wz}
\begin{eqnarray}\label{gamma2}
{d\Gamma\over d\cos\theta} =
\int_t
\frac{\beta_t^{1/2}\lambda^{1/2}_t}{(8\pi m_B)^3}|\bar {\cal A}|^2\;dt\;,
\end{eqnarray}
where  $t\equiv m_{12}^2$, $\beta_t=1-(m_{\bf B}+m_{\bf \bar B'})^2/t$,
$\lambda_t=[(m_B+m_M)^2+t][(m_B-m_M)^2+t]$,
and  $\theta$ is the angle between the moving directions of $\bf B$ and $M$.

\section{Numerical analysis}
\begin{table}[t!]
\caption{The values of $\alpha_i$ and $\beta_i$ with $N_c=2,\,3$, and $\infty$.}\label{alpha_i}
\begin{tabular}{|c|ccc|}
\hline
$\alpha_i\,(\beta_i)$ & $N_c=2$ &  $N_c=3$ & $N_c=\infty$ \\\hline
$10^4\alpha_1$        & $-14.6-11.0i$ &  $-15.4- 11.6i$ & $-16.9 - 13.0i$ \\
$10^4 \alpha_2^+$                  & $-17.2 -4.4i$ &  $-1.5-0.3i$ & $29.8+8.0i$ \\
$10^4 \alpha_2^-(-\beta_2)$   & $12.2 - 1.7i$ &  $9.6-0.2i$ & $4.5+2.8i$ \\
$10^4\alpha_3^+$  & $-22.0-4.5i$ &  $-15.8-3.4i$ & $-3.4-1.2i$ \\
$10^4\alpha_3^-(-\beta_3)$  & $-7.3-3.3i$ &  $-10.3-3.5i$ & $-16.2-3.9i$ \\
$10^4\alpha_6$ & $48.1+6.7i$ &  $50.1+7.1i$ & $54.3+7.9i$ \\
$10^4\alpha_6^s(\beta_6)$ & $48.6+6.5i$ &  $50.7+7.0i$ & $55.0+7.9i$ \\
\hline
\end{tabular}
\end{table}

In the numerical analysis, we use the Wolfenstein parameters
for the CKM matrix elements:
\begin{eqnarray}
&&V_{ub}=A\lambda^3(\rho-i\eta),\,V_{tb}=1\,,\nonumber\\
&&V_{us}=\lambda,\,V_{ts}=-A\lambda^2,
\end{eqnarray}
with $\lambda$, $A$, $\rho=\bar \rho/(1-\lambda^2/2)$ 
and $\eta=\bar \eta/(1-\lambda^2/2)$, given by~\cite{pdg}
\begin{eqnarray}\label{4_number}
\lambda=0.22453\pm 0.00044\,,
A=0.836\pm 0.015\,,
\bar \rho=0.122^{+0.018}_{-0.017}\,,
\bar \eta=0.355^{+0.012}_{-0.011}\,.
\end{eqnarray}
In the adoption of the effective Wilson coefficients $c_i^{eff}$ in Ref.~\cite{ali},
the values of $\alpha_i$ and $\beta_i$ 
in Eqs.~(\ref{amp1}), (\ref{amp1b}) and (\ref{alpha_beta}) are given in Table~\ref{alpha_i}
with $N_c=(2,3,\infty)$ to estimate the non-factorizable effects.
For the $0\to (\eta,\eta^\prime)$ productions and 
$B\to (\eta,\eta^\prime)$ transitions, 
one gets~\cite{Beneke:2002jn,MFD,Fan:2012kn}
\begin{eqnarray}\label{para}
&&
(f^n_{\eta},f^n_{\eta^\prime},f^s_{\eta},f^s_{\eta^\prime})=
(0.108,\,0.089\,,-0.111,\,0.136)\,\text{GeV}\,,\nonumber\\
&&
(h^s_{\eta},h^s_{\eta^\prime})=(-0.055,\,0.068)\,\text{GeV}^3\,,\nonumber\\
&&
(F^{B\eta_n},\sigma_{11},\sigma_{12},\sigma_{01},\sigma_{02})
=(0.33,\,0.48,\,0,\,0.76,\,0.28)\,,\nonumber\\
&&
(F^{B_s\eta_s},\sigma_{11},\sigma_{12},\sigma_{01},\sigma_{02})
=(0.36,\,0.60,\,0.20,\,0.80,\,0.40)\,,
\end{eqnarray}
with $M_V=5.32$~GeV, resulting in
$(F^{B\eta},F^{B\eta'})=(0.26,0.21)$ and $(F^{B_s\eta},F^{B_s\eta'})=(-0.23,0.28)$ 
by Eq.~(\ref{form2b}).

To extract the $0\to {\bf B\bar B'}$ baryonic form factors,
the minimal $\chi^2$ fitting method has been used to fit with 20 data points, where
11 of them are from the branching ratios of 
$D_s^+\to p\bar n$, $\bar B^0_{(s)}\to p\bar p$, $B^-\to\Lambda\bar p$,
$\bar B^0\to n\bar p D^{*+}(\Lambda\bar p D^{(*)+})$,
$\bar B^0(B^-)\to\Lambda\bar p\pi^{+(0)}$, 
$B^-\to\Lambda\bar p \rho^0$ and $B^-\to \Lambda\bar \Lambda K^-$,
4  the angular distribution asymmetries of 
$\bar B^0\to \Lambda\bar p D^{(*)+}$, $\bar B^0(B^-)\to\Lambda\bar p\pi^{+(0)}$
and 5  the angular distribution in $\bar B^0\to \Lambda\bar p\pi^+$~\cite{Wang:2007as}.
This presents a reasonable fit with $\chi^2/d.o.f\simeq 2.3$,
where $d.o.f$ stands for the degree of freedom.
Hence, we adopt the fitted values to be~\cite{Hsiao:2016amt,Hsiao:2017nga,Geng:2016fdw}
\begin{eqnarray}\label{fitC1}
&&(C_{||},\,\delta C_{||})=(154.4\pm 12.1,\,19.3\pm 21.6)\;{\rm GeV}^{4}\,,\nonumber\\
&&(C_{\overline{||}},\,\delta C_{\overline{||}})=(18.1\pm 72.2,\,-477.4\pm 99.0)\;{\rm GeV}^{4}\,,\nonumber\\
&&(\bar C_{||},\,\delta\bar C_{||})=(537.6\pm 28.7,\,-342.3\pm 61.4)\;{\rm GeV}^{4}\,.
\end{eqnarray}
Here, we have assumed that the timelike baryonic form factors
are real. In general, they can be complex numbers 
if some resonances are involved with un-calculable strong phases.
However, these  phases are believed to be negligible.
For the $B\to {\bf B\bar B'}$ transition ones,
the extractions depend on 28 data points with 
7 from the branching ratios of 
$B^-\to p\bar p (K^-,\pi^-)$, $B^-\to p\bar p e^-\bar \nu_e$ and
$\bar B^0\to p\bar p (K^{(*)0},D^{(*)0})$,
3  the $CP$ violating asymmetries of 
$B^-\to p\bar p (K^{(*)-},\pi^-)$ and
2  the angular distribution asymmetries of 
$B^-\to p\bar p (K^-,\pi^-)$,
together with 16 data points from the angular distributions in
$B^-\to p\bar p (K^-,\pi^-)$~\cite{Wei:2007fg},
resulting in $\chi^2/d.o.f\simeq 0.8$ for a reasonable fit also. 
The values of $D_i$ are given by~\cite{Hsiao:2016amt,Geng:2016fdw}
\begin{eqnarray}\label{fitD2}
&&
D_{||}=(45.7\pm 33.8)\;{\rm GeV}^{5}\,,
(D_{||}^4,D_{||}^5)=(6.5\pm 18.1,-147.1\pm 29.3)\;{\rm GeV}^{4}\,,\nonumber\\
&&
(\bar D_{||},\bar D_{||}^2,\bar D_{||}^3)=(35.2\pm 4.8,-22.3\pm 10.2, 504.5\pm 32.4)\;{\rm GeV}^{4}\,.
\end{eqnarray}
With the theoretical inputs in Eqs.~(\ref{fitC1}) and (\ref{fitD2}), one has well explained
the observations of 
${\cal B}(\bar B^0_s\to \bar p\Lambda K^+ +p\bar \Lambda K^-)$ and
${\cal B}(B\to p\bar p MM)$~\cite{Geng:2016fdw,Hsiao:2017nga}.

\begin{table}[t!]
\caption{Numerical results for 
${\cal B}(B^-\to \Lambda\bar p\eta^{(\prime)})$ and
${\cal B}(\bar B^0_s\to \Lambda\bar \Lambda \eta^{(\prime)})$,
with ${\cal B}_\pm={\cal B}_1+{\cal B}_2\pm {\cal B}_{1\cdot 2}$, 
where $({\cal B}_1, {\cal B}_2, {\cal B}_{1\cdot 2})$ are
denoted as the partial branching ratios from the amplitudes
${\cal A}_1$, ${\cal A}_2$ and the interferences,
while the errors come from the non-factorizable effects and form factors,
respectively.}\label{tab1}
\begin{tabular}{|c|c|c|}
\hline
branching ratios & ${\cal B}_+$&${\cal B}_-$\\\hline
$10^{6}{\cal B}(B^-\to \Lambda\bar p\eta)$                
&$5.3\pm 0.7\pm 1.2$ &$4.0\pm 0.6\pm 0.4$\\
$10^{6}{\cal B}(B^-\to \Lambda\bar p\eta')$               
&$3.3\pm 0.6\pm 0.4$ &$4.6\pm 0.7\pm 0.9$\\
$10^{6}{\cal B}(\bar B^0_s\to \Lambda\bar \Lambda \eta)$    
&$1.2\pm 0.2\pm 0.2$&$2.1\pm 0.4 \pm 0.5$\\
$10^{6}{\cal B}(\bar B^0_s\to \Lambda\bar \Lambda \eta')$   
&$2.6\pm 0.5\pm 0.6$ &$1.5\pm 0.1\pm 0.4$\\
\hline
\end{tabular}
\end{table}
\begin{figure}[t!]
\centering
\includegraphics[width=2.4in]{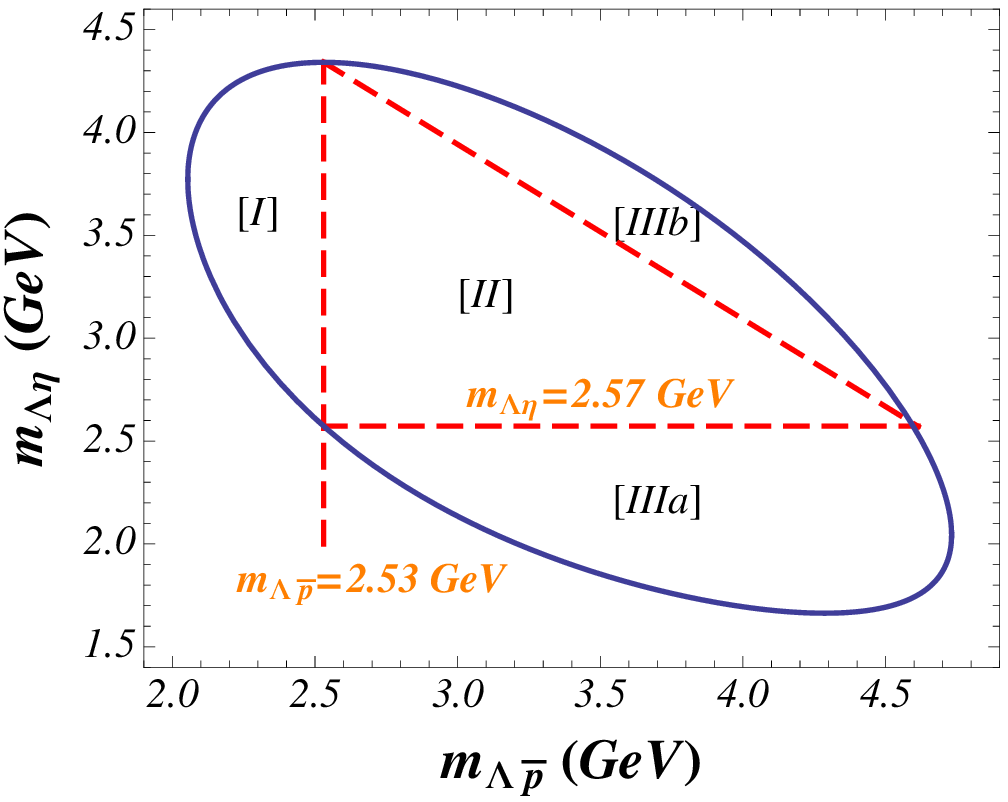}
\includegraphics[width=2.75in]{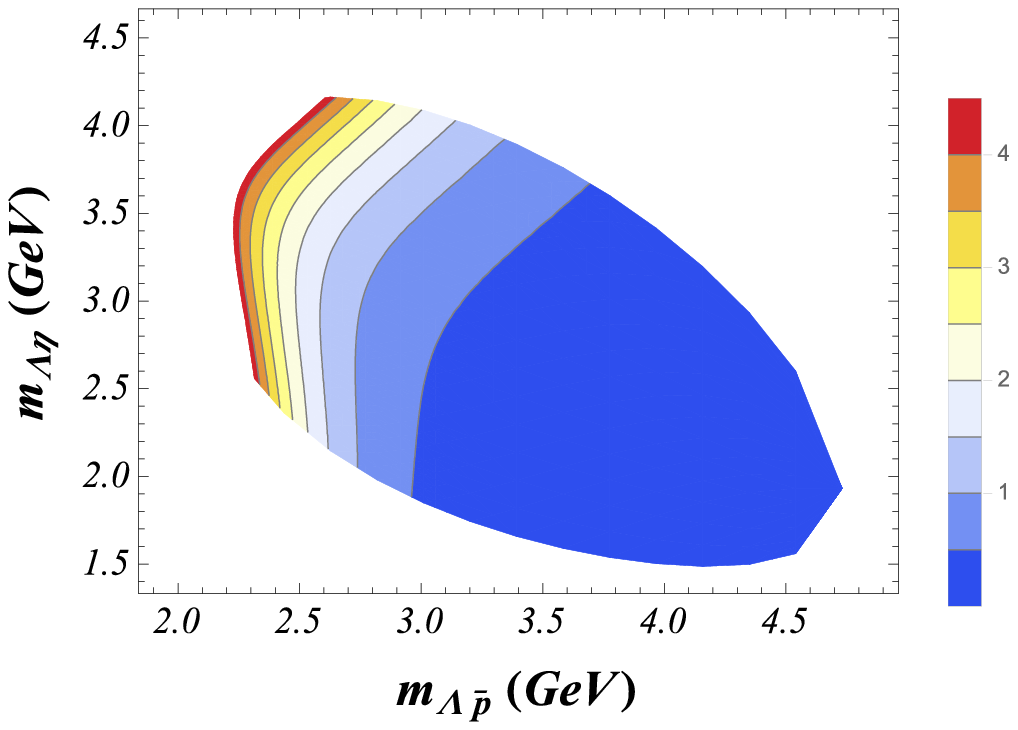}
\caption{
The kinematical allowed regions of I, II, IIIa and IIIb
(left panel) and Dalitz plot distribution (right panel)
 in the plane of 
 $m_{\Lambda\bar p}$ and $m_{\Lambda\eta}$
for $B^-\to\Lambda\bar p\eta$.}\label{DP}
\end{figure}

Since the $0\to{\bf B\bar B'}$  and $B\to {\bf B\bar B'}$ baryonic
transition form factors are separately extracted from the data, 
it is possible to have overall positive or negative signs
between $C$ and $D$ in Eqs.~(\ref{fitC1}) and (\ref{fitD2}),  
causing two different scenarios for the interferences.
In Table~\ref{tab1},
we present the results for 
${\cal B}(B^-\to \Lambda\bar p\eta^{(\prime)})$ and
${\cal B}(\bar B^0_s\to \Lambda\bar \Lambda\eta^{(\prime)})$
with ${\cal B}_{\pm}\equiv{\cal B}_1+{\cal B}_2\pm {\cal B}_{1\cdot 2}$,
where the notations of ``$\pm$'' are due to the undetermined relative signs, and
$({\cal B}_1, {\cal B}_2, {\cal B}_{1\cdot 2})$ are
denoted as the partial branching ratios from the amplitudes
${\cal A}_1$, ${\cal A}_2$ and the interferences, respectively.
Note that the errors in Table~\ref{tab1} arise from
the estimations of the non-factorizable effects in the generalized factorization
with $N_c=2,3,\infty$ for the parameters in Table~\ref{alpha_i}, and 
the uncertainties in the form factors of the $0\to{\bf B\bar B'}$ productions
and $B\to{\bf B\bar B'}$ transitions in Eqs.~(\ref{fitC1}) and (\ref{fitD2}). 
On the other hand, 
the uncertainties from the CKM matrix elements in Eq.~(\ref{4_number})
have been computed to be negligibly small.

In Table~\ref{tab1}, we have used the central values of
 $({\cal B}_1,{\cal B}_2,{\cal B}_{1\cdot 2})= (2.92,1.73,0.65)\times 10^{-6}$
 and 
$({\cal B}'_1,{\cal B}'_2,{\cal B}'_{1\cdot 2})= (1.71,2.24,-0.61)\times 10^{-6}$
 for $B^-\to \Lambda\bar p\eta^{(\prime)}$.
Clearly, the results of  $|{\cal B}_{1\cdot 2}^{(\prime)}|\sim {\it O}(10^{-6})$
indicate  sizable  interferences.
As shown in the table, we find that 
${\cal B}_{1\cdot 2}^{(\prime)}$
causes a constructive (destructive) interfering effect in 
${\cal B}_+(B^-\to \Lambda\bar p\eta^{(\prime)})$, and 
a destructive (constructive) interfering one in 
${\cal B}_-(B^-\to \Lambda\bar p\eta^{(\prime)})$.
Besides, the inequalities of 
${\cal B}_1>{\cal B}'_1$ and  
${\cal B}_2<{\cal B}'_2$ are due to 
$F^{B\eta}>F^{B\eta'}$ 
and $|h^s_{\eta}|<|h^s_{\eta^\prime}|$,
respectively.
Similarly, one has that
$({\cal B}_{s1},{\cal B}_{s2},{\cal B}_{s1\cdot s2})=(1.33,0.33,-0.46)\times 10^{-6}$
for $\bar B^0_s\to \Lambda\bar \Lambda \eta$
and 
$({\cal B}'_{s1},{\cal B}'_{s2},{\cal B}'_{s1\cdot s2})=(1.73,0.29,0.57)\times 10^{-6}$
for $\bar B^0_s\to \Lambda\bar \Lambda \eta^{\prime}$,
which present that 
${\cal B}_{s1\cdot s2}^{(\prime)}$
has a destructive (constructive) interfering effect in
${\cal B}_+(\bar B^0_s\to \Lambda\bar \Lambda\eta^{(\prime)})$, and 
a constructive (destructive) interfering one in 
${\cal B}_-(\bar B^0_s\to \Lambda\bar \Lambda\eta^{(\prime)})$.
As a result, in terms of
$|{\cal B}_{(s)1\cdot (s)2}^{(\prime)}|\sim {\it O}(10^{-6})$
being traced back to the interferences 
between the two decaying configurations in Fig.~\ref{dia},
we conclude that $B^-\to \Lambda\bar p\eta^{(\prime)}$ and
$\bar B^0_s\to \Lambda\bar \Lambda \eta^{(\prime)}$
are like $B\to K^{(*)}\eta^{(\prime)}$
to have large values from the interferences,
in comparison with 
${\cal B}(\Lambda_b\to \Lambda\eta)\simeq 
{\cal B}(\Lambda_b\to \Lambda\eta')$~\cite{Ahmady:2003jz,Geng:2016gul}
and ${\cal B}(\Lambda_c^+\to p\eta)\simeq 
{\cal B}(\Lambda_c^+\to p\eta')$~\cite{Geng:2017esc,Geng:2018plk},
which show less important interferences.

By following Refs.~\cite{Krankl:2015fha,Cheng:2016shb},
we present the kinematical allowed regions 
and Dalitz plot distribution in the plane of 
$m_{\Lambda\bar p}$ and $m_{\Lambda\eta}$ for $B^-\to\Lambda\bar p\eta$
in Fig.~\ref{DP} to illustrate 
the generic features in $B\to{\bf B\bar B'}M$.
As shown in the left panel in the figure,
the allowed area  can be divided into four different regions, denoted as I, II, IIIa and IIIb, respectively.
In Region~I, 
${\bf B}$ and ${\bf \bar B'}$ can move collinearly, 
with the recoiled $M$ in the opposite direction.
In Region~II,
$\bf B$, $\bf \bar B'$ and $M$ all have large energies, so that 
none of any two final states can be back-to-back.
In Region IIIa(b), $M$ and ${\bf B}({\bf \bar B'}$) move collinearly, 
with $\bf \bar B'$ ($\bf B$) being energetic 
and separated from the meson-(anti)baryon system. 
Since the collinear moving di-baryon in Region~I
and meson-(anti)baryon in Region~IIIa(b)
cause different kinds of quasi-two-body decays, 
the $t$ and $s$ ($u$)-channel contributions
should be dominant, respectively, where
$t\equiv (p_{\bf B}+p_{\bf \bar B'})^2$, 
$s\equiv (p_{\bf B}+p_{M})^2$ and $u\equiv (p_{\bf \bar B'}+p_M)^2$
are the Mandelstam variables. 
In Region~II, 
the three channels are supposed to contribute with $(s,t,u)\sim m_B^2/3$.

\begin{figure}[t!]
\centering
\includegraphics[width=2.4in]{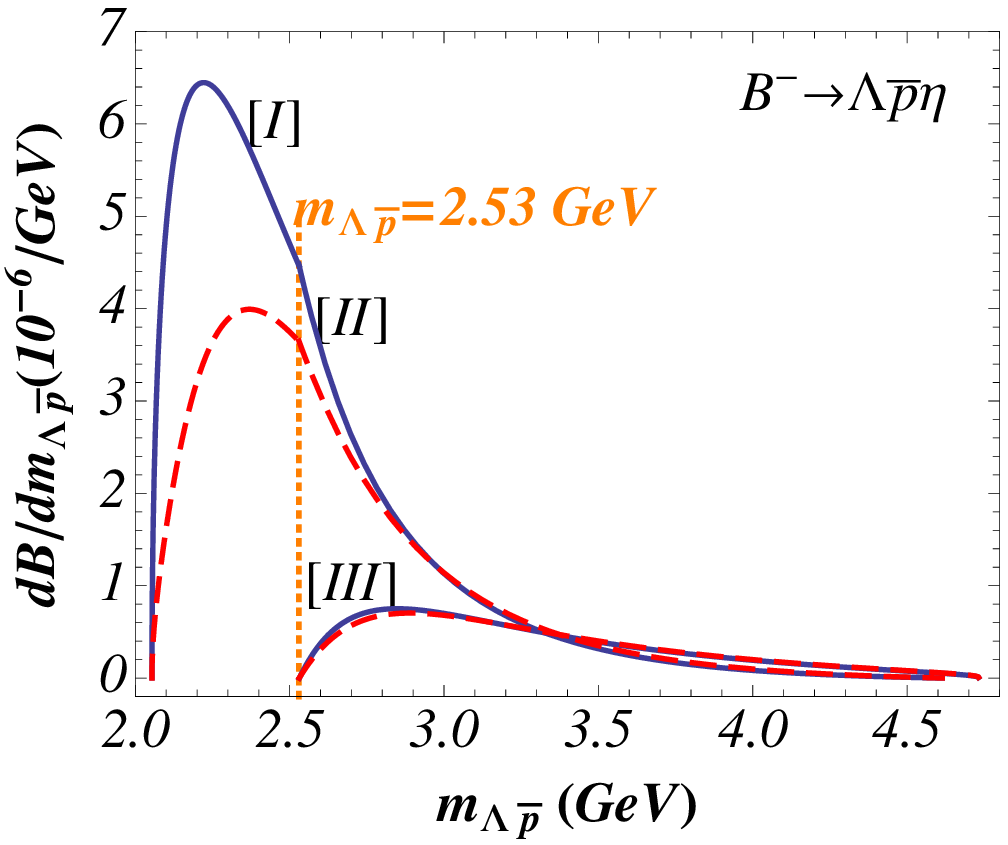}
\includegraphics[width=2.4in]{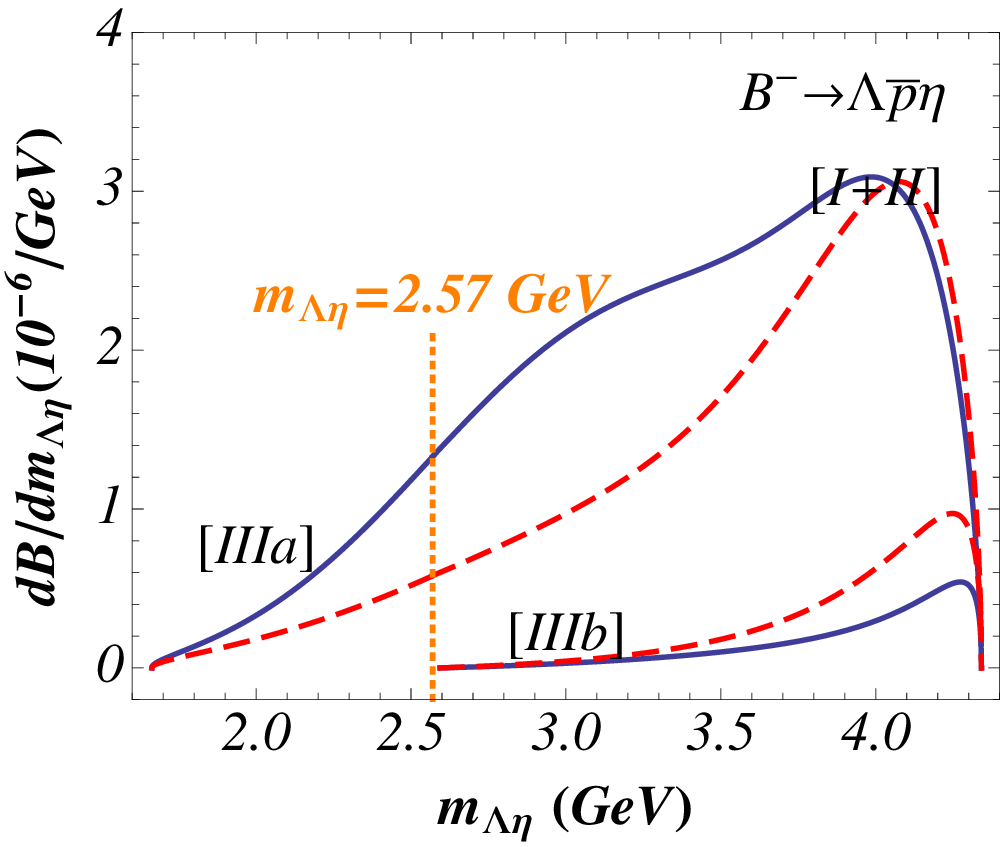}
\includegraphics[width=2.4in]{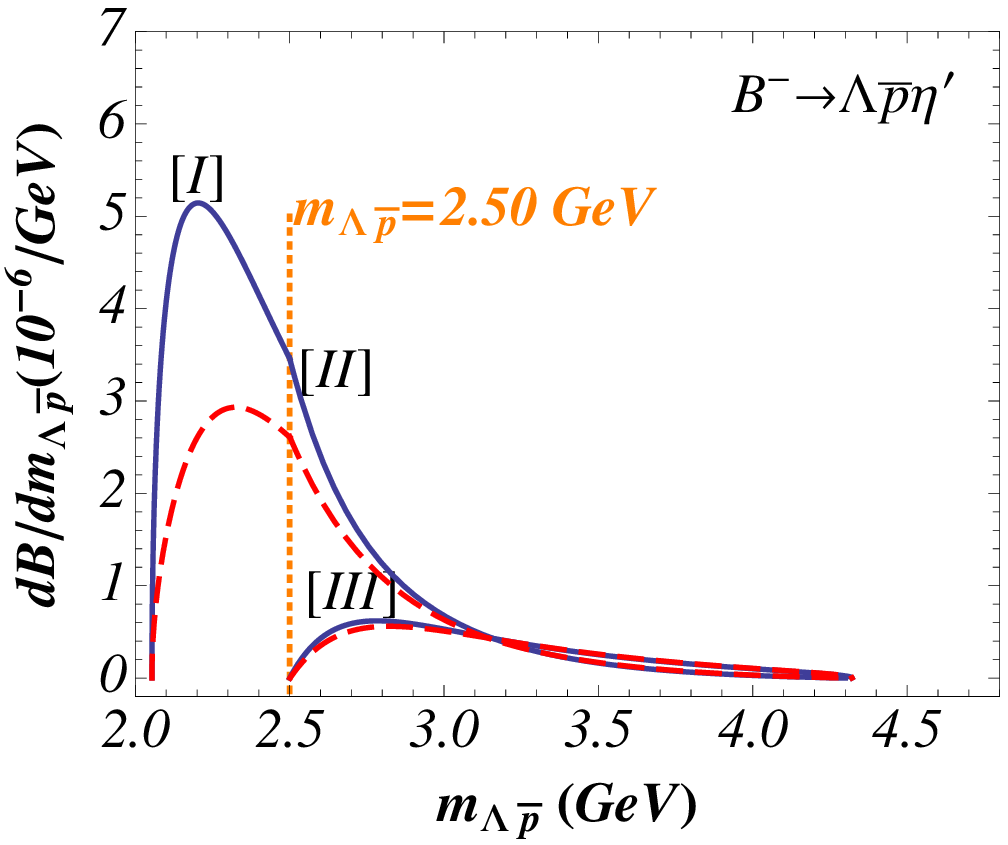}
\includegraphics[width=2.4in]{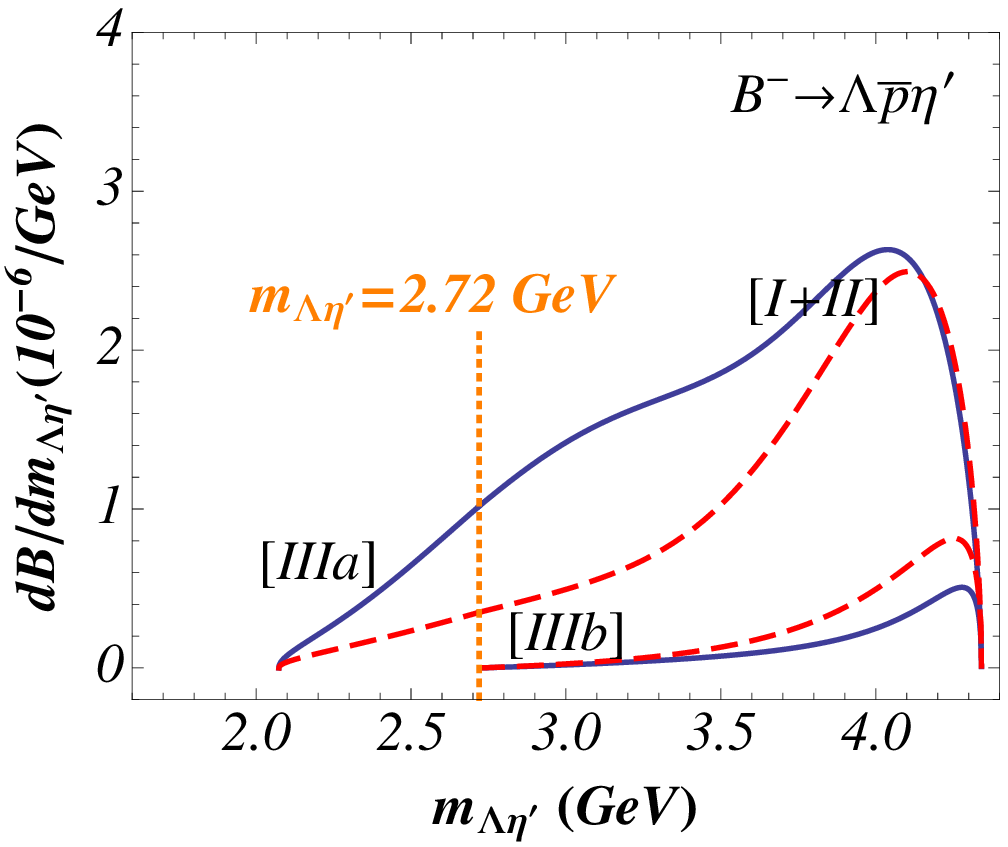}
\includegraphics[width=2.4in]{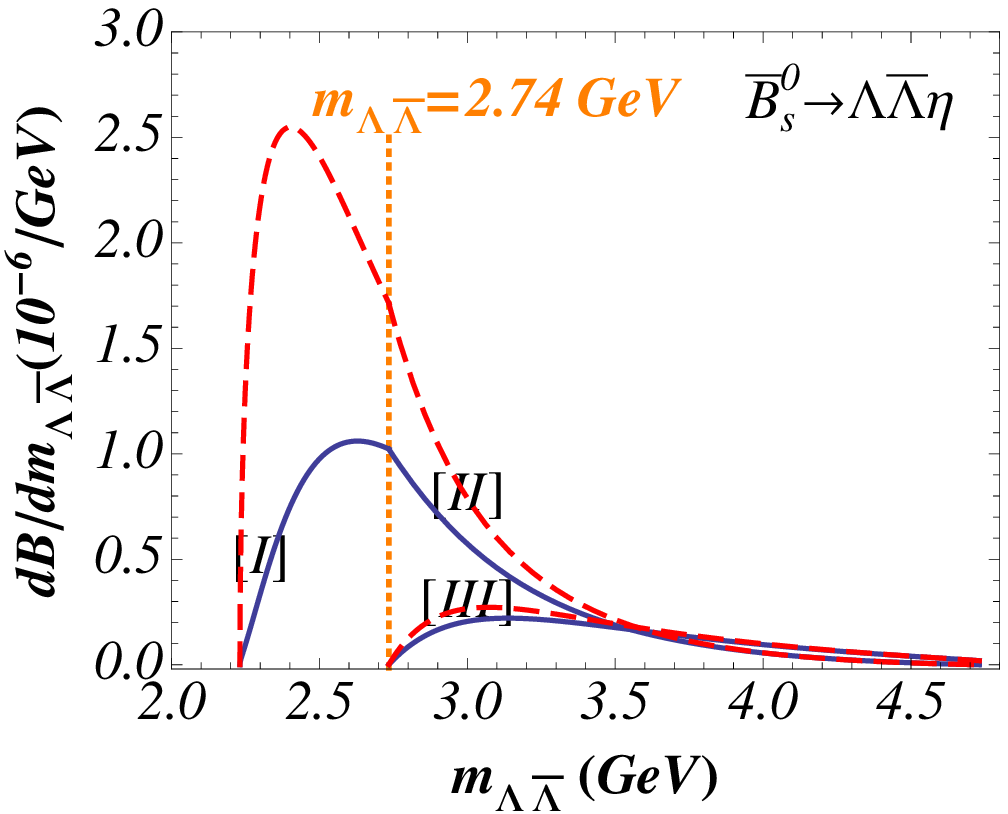}
\includegraphics[width=2.4in]{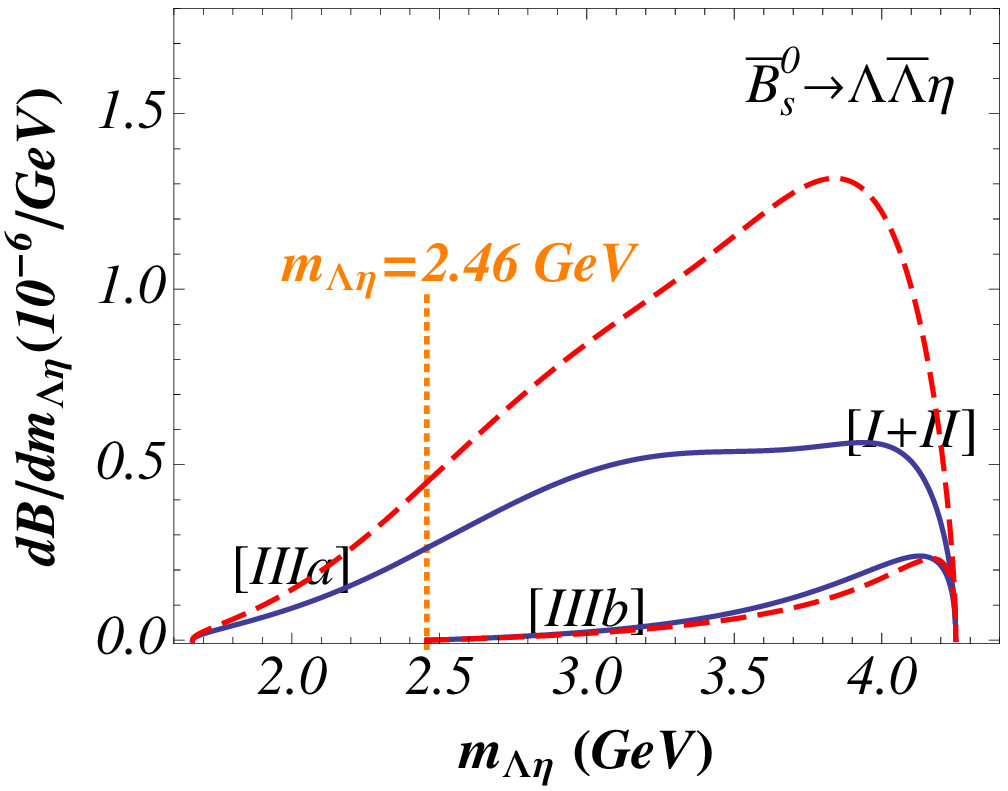}
\includegraphics[width=2.4in]{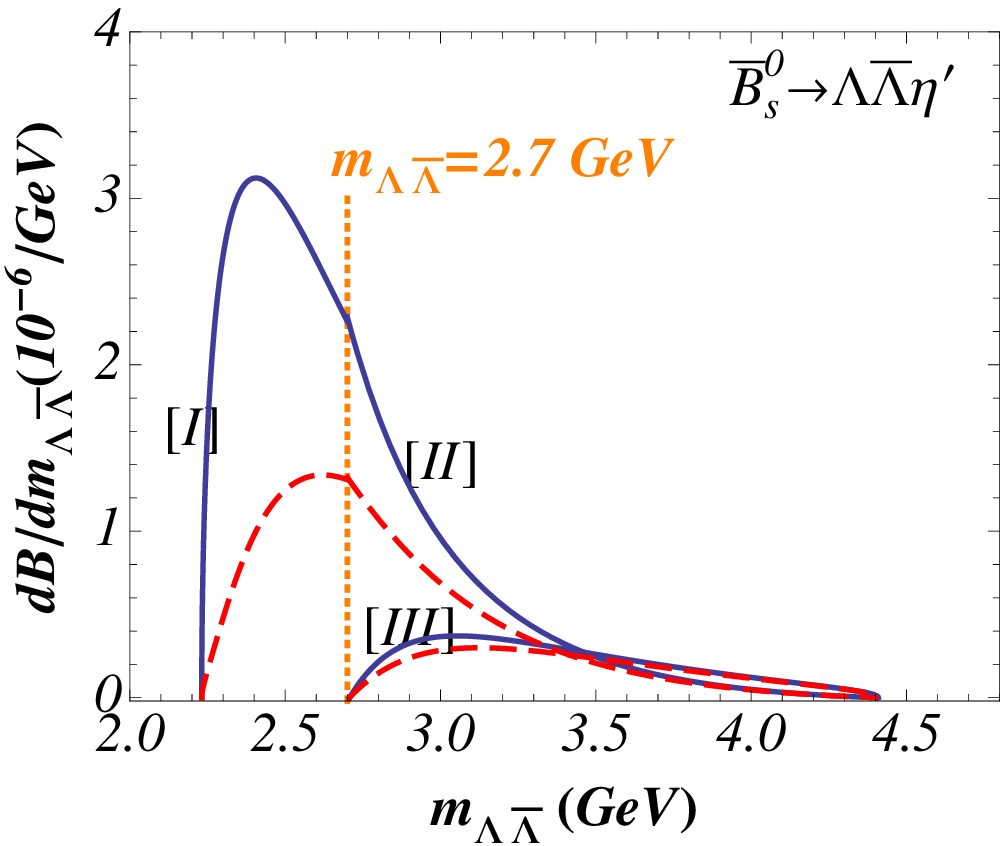}
\includegraphics[width=2.4in]{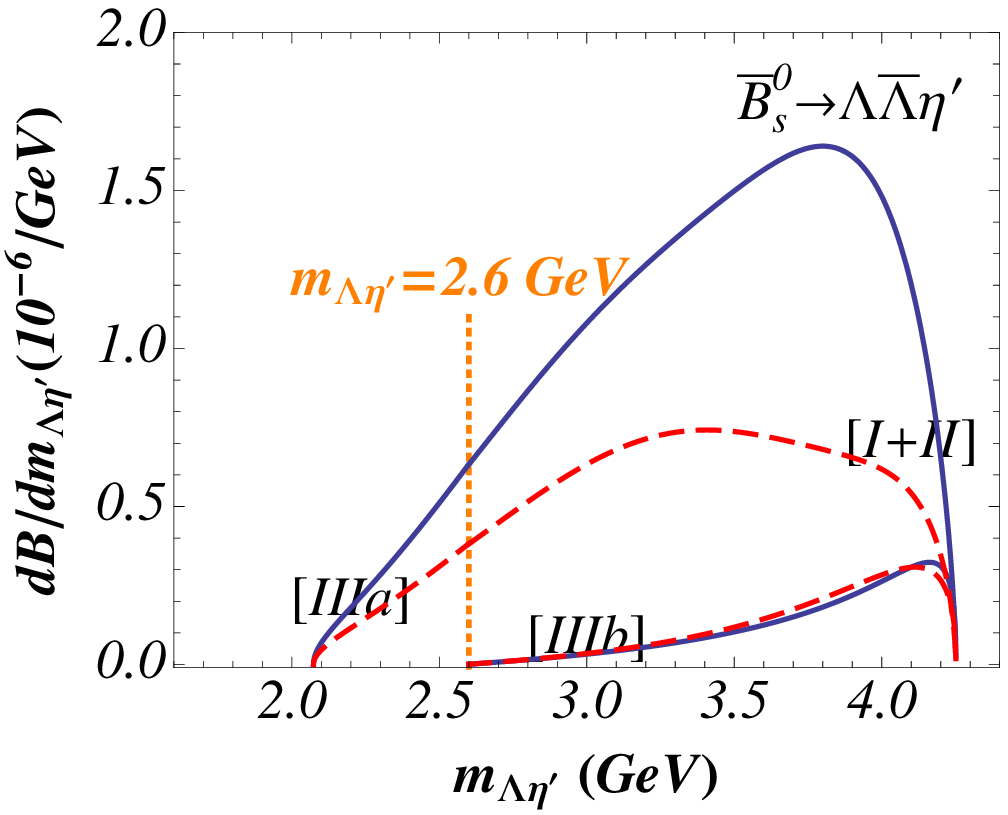}
\caption{Decay spectra versus $m_{\bf B\bar B'}$ (left) and $m_{{\bf B}\eta^{(\prime)}}$ (right)
of the three-body $B\to {\bf B\bar B'}\eta^{(\prime)}$ decays,
where the solid (dash) curves for $B^-\to \Lambda\bar p\eta$ 
with the constructive (destructive) interfering effects
correspond to the kinematical regions in the left panel of Fig.~\ref{DP},
while those for $B^-\to \Lambda\bar p\eta^\prime$ and 
$\bar B^0_s\to \Lambda\bar \Lambda\eta^{(\prime)}$ 
are similarly presented.}\label{spectra}
\end{figure}

\begin{figure}[t!]
\centering
\includegraphics[width=2.2in]{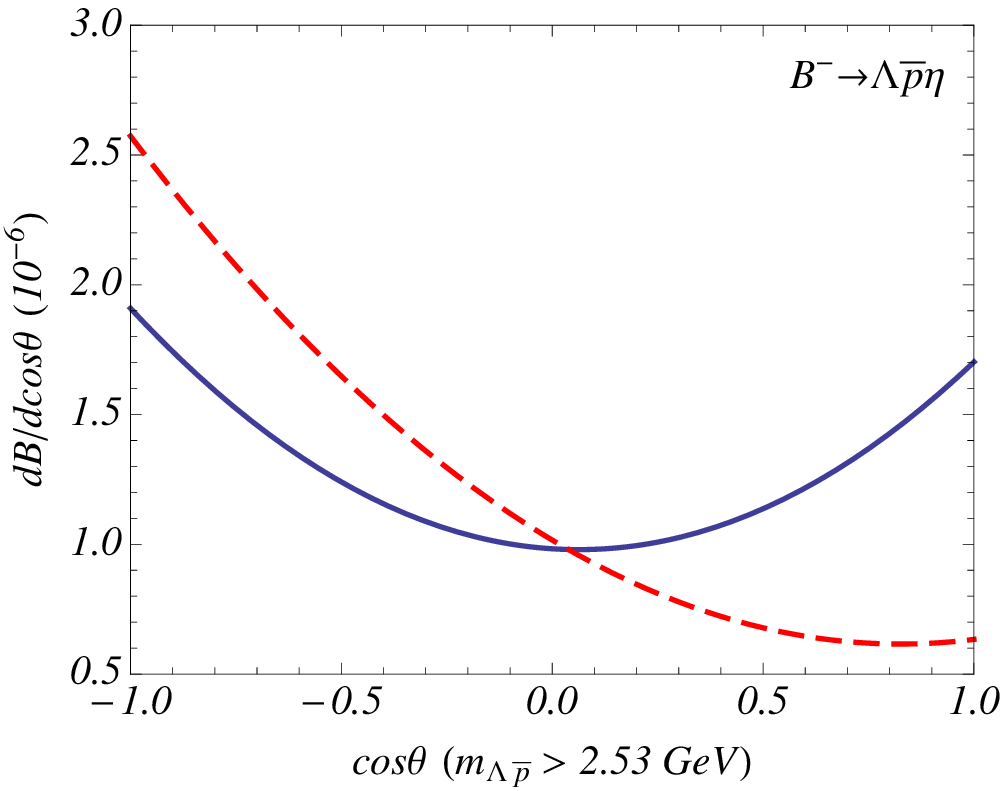}
\includegraphics[width=2.2in]{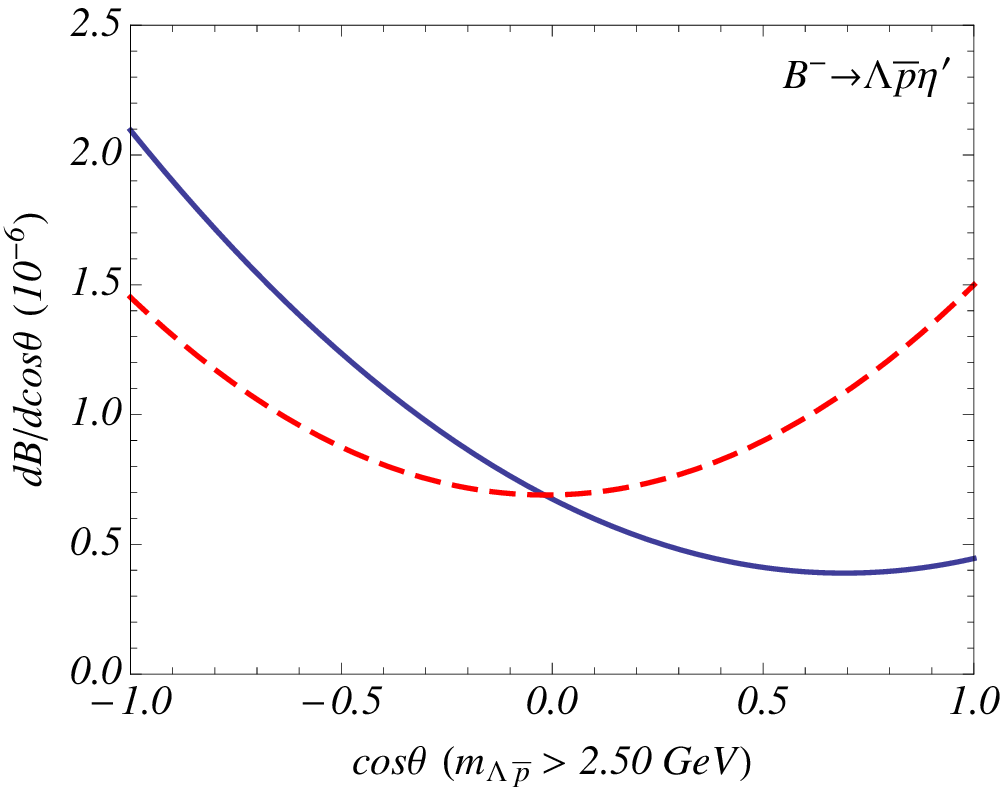}
\includegraphics[width=2.2in]{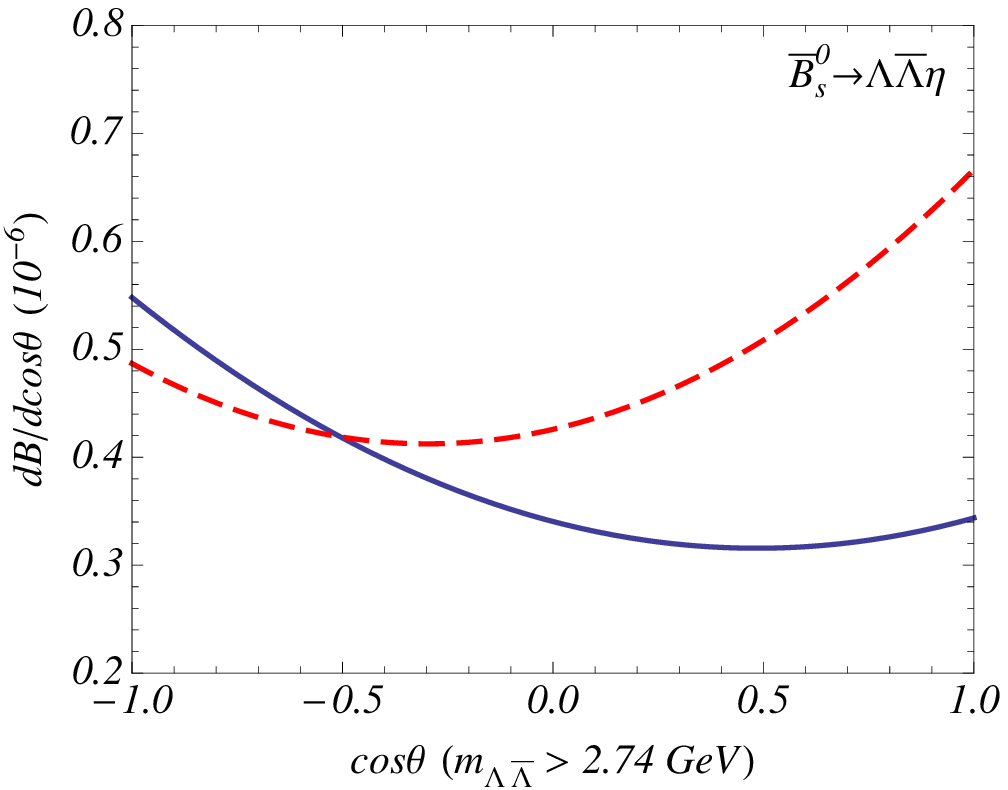}
\includegraphics[width=2.2in]{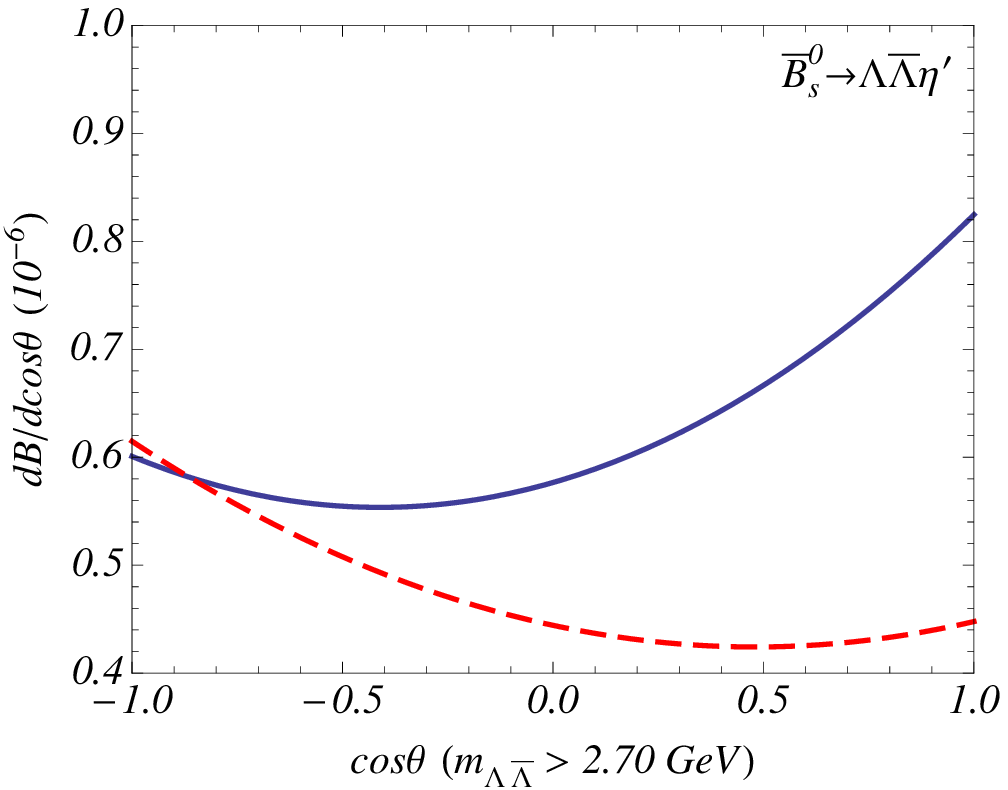}
\caption{Angular distributions of $B\to{\bf B\bar B'}\eta^{(\prime)}$ versus $\cos\theta$ 
with  $\theta$ being the angle between the baryon and meson moving directions, 
where the solid (dash) curves correspond to  the constructive (destructive) interference  effects.
}\label{AD}
\end{figure}

Although Regions I, II and IIIa(b) have distinct dynamic properties, 
we  assume that
the expressions of Eqs.~(\ref{amp1}) and (\ref{amp1b})
for the di-baryon threshold effect in Region~I
can be extended to the other regions.
In fact, the extension has been demonstrated to be able to
describe the $\bar B^0\to p\bar p D^0$ spectra
at different energy ranges~\cite{R1,R2,R_BABAR},
where the data points for the spectrum vs. $m_{Dp}$
are measured at the range of $m_{p\bar p}>2.29$ GeV, 
which correspond to the regions II and III of Fig.~\ref{DP}.
In order that the extension of the amplitudes in Eqs.~(\ref{amp1}) and (\ref{amp1b})
can be tested by the future observations, we present
the spectra versus $m_{\bf B\bar B'}$ and $m_{{\bf B}\eta^{(\prime)}}$
in the three-body $B\to {\bf B\bar B'}\eta^{(\prime)}$ decays in Fig.~\ref{spectra} 
for the different kinematic regions in the Dalitz plots.
Besides,
we show the angular distributions with $m_{\bf B\bar B'}>2.5-2.7$ GeV in Fig.~\ref{AD},
to be compared to the future measurements.
Note that
$\cos\theta\simeq 0$ with $\theta\simeq 90^\circ$
corresponds to the central area of Region~II for the Dalitz plots.
%
%

Like the $B\to K^{(*)}\eta^{(\prime)}$ decays,
it is possible that 
$B\to{\bf B\bar B'}\eta^{(\prime)}$
can help to improve the knowledge of the underlying QCD anomaly
for the $\eta-\eta'$ mixing. 
Provided that the decays of $B\to {\bf B\bar B'}\eta^{(\prime)}$ are well measured,
the experimental values to be inconsistent with the theoretical calculations
will hint at some possible additional effects to the $\eta-\eta'$ mixing,
such as the $\eta$-$\eta'$-$G$ mixing
with $G$ denoting the pseudoscalar glueball state~\cite{Cheng:2008ss}.
Moreover,  
the gluonic contributions 
to the $B(\bar B_s^0)\to \eta^{(\prime)}$ transition form factors~\cite{Duplancic:2015zna}
could also lead to visible effects.

\section{Conclusions}
We have studied the three-body baryonic $B$ decays of
$B^-\to \Lambda\bar p\eta^{(\prime)}$ and 
$\bar B^0_s\to \Lambda\bar \Lambda \eta^{(\prime)}$.
Due to the interference effects between 
 $b\to s n\bar n\to s\eta_n$ and $b\to s\bar s s\to s\eta_s$,
which can be constructive or destructive, we have predicted that 
${\cal B}(B^-\to\Lambda\bar p\eta,\Lambda\bar p\eta')=(5.3\pm 1.4,3.3\pm 0.7)\times 10^{-6}$
or $(4.0\pm 0.7,4.6\pm 1.1)\times 10^{-6}$, to be compared to the searching results 
by LHCb and BELLE. 
We have also found that
${\cal B}(\bar B^0_s\to \Lambda\bar \Lambda \eta,\Lambda\bar \Lambda \eta')
=(1.2\pm 0.3,2.6\pm 0.8)\times 10^{-6}$ or $(2.1\pm 0.6,1.5\pm 0.4)\times 10^{-6}$.
In our calculations, the errors came from 
the estimations of the non-factorizable effects in the generalized factorization,
together with the uncertainties from the form factors of the $0\to {\bf B\bar B'}$ productions
and $B\to {\bf B\bar B'}$ transitions, 
which are due to the fit with the existing data for the baryonic $B$ decays.
Due to the fact that the contributions from  Region~II and 
the resonant meson-baryon pairs in Regions III in Fig.~\ref{DP}
are not considered properly,
our results just provide an order of magnitude estimation on branching ratios.

\section*{ACKNOWLEDGMENTS}
We would like to thank Dr. Minzu Wang for useful discussions.
This work was supported in part by National Science Foundation of China (11675030),
National Center for Theoretical Sciences, and
MoST (MoST-104-2112-M-007-003-MY3 and MoST-107-2119-M-007-013-MY3).


\begin{thebibliography}{99}
\bibitem{FKS}
T.~Feldmann, P.~Kroll and B.~Stech,
Phys.\ Rev.\ D {\bf 58}, 114006 (1998); Phys.\ Lett.\ B {\bf 449}, 339 (1999).

\bibitem{pdg}
M.~Tanabashi {\it et al.} [Particle Data Group],
Phys.\ Rev.\ D {\bf 98}, 030001 (2018).
  

\bibitem{Hsiao:2015iiu} 
Y.K.~Hsiao, C.F.~Chang and X.G.~He,
Phys.\ Rev.\ D {\bf 93}, 114002 (2016).


\bibitem{Aaij:2015qga} 
R.~Aaij {\it et al.} [LHCb Collaboration],
Phys.\ Rev.\ Lett.\  {\bf 115}, 051801 (2015).

\bibitem{Ahmady:2003jz} 
M.R.~Ahmady, C.S.~Kim, S.~Oh and C.~Yu,
Phys.\ Lett.\ B {\bf 598}, 203 (2004).

\bibitem{Geng:2016gul} 
C.Q.~Geng, Y.K.~Hsiao, Y.H.~Lin and Y.~Yu,
Eur.\ Phys.\ J.\ C {\bf 76}, 399 (2016).

\bibitem{LbtoLeta}
R.~Aaij {\it et al.} [LHCb Collaboration], JHEP {\bf 1509}, 006 (2015).

\bibitem{Geng:2017esc} 
C.Q.~Geng, Y.K.~Hsiao, Y.H.~Lin and L.L. Liu,
Phys.\ Lett.\ B {\bf 776}, 265 (2018). 

\bibitem{Geng:2018plk} 
C.Q.~Geng, Y.K.~Hsiao, C.W.~Liu and T.H.~Tsai,
JHEP {\bf 1711}, 147 (2017);
Phys.\ Rev.\ D {\bf 97}, 073006 (2018). 

\bibitem{Hou:2000bz} 
W.S.~Hou and A.~Soni, Phys.\ Rev.\ Lett.\  {\bf 86}, 4247 (2001).

\bibitem{Chua:2002wn} 
C.K.~Chua, W.S.~Hou and S.Y.~Tsai, Phys.\ Rev.\ D {\bf 66}, 054004 (2002).

\bibitem{Chua:2002yd} 
C.K.~Chua and W.S.~Hou, Eur.\ Phys.\ J.\ C {\bf 29}, 27 (2003).

\bibitem{Geng:2005wt} 
C.Q.~Geng and Y.K.~Hsiao, Phys.\ Rev.\ D {\bf 72}, 037901 (2005);
Int.\ J.\ Mod.\ Phys.\ A {\bf 21}, 897 (2006).

\bibitem{Geng:2005fh} 
C.Q.~Geng and Y.K.~Hsiao,
Phys.\ Lett.\ B {\bf 619}, 305 (2005).

\bibitem{Geng:2006wz} 
C.Q.~Geng and Y.K.~Hsiao,
Phys.\ Rev.\ D {\bf 74}, 094023 (2006).

\bibitem{Geng:2006jt} 
C.Q.~Geng, Y.K.~Hsiao and J.N.~Ng, 
Phys.\ Rev.\ Lett.\  {\bf 98}, 011801 (2007).

\bibitem{Chen:2008sw} 
C.H.~Chen, H.Y.~Cheng, C.Q.~Geng and Y.K.~Hsiao,
Phys.\ Rev.\ D {\bf 78}, 054016 (2008).

\bibitem{Geng:2011pw} 
C.~Q.~Geng and Y.~K.~Hsiao,
Phys.\ Rev.\ D {\bf 85}, 017501 (2012).

\bibitem{Hsiao:2016amt} 
Y.K.~Hsiao and C.Q.~Geng,
Phys.\ Rev.\ D {\bf 93}, 034036 (2016).

\bibitem{Geng:2016fdw} 
C.Q.~Geng, Y.K.~Hsiao and E.~Rodrigues,
Phys.\ Lett.\ B {\bf 767}, 205 (2017).

\bibitem{Hsiao:2017nga} 
Y.K.~Hsiao and C.Q.~Geng,
Phys.\ Lett.\ B {\bf 770}, 348 (2017).

\bibitem{Lu:2018qbw} 
P.C.~Lu {\it et al.},
arXiv:1807.10503 [hep-ex].

\bibitem{Abe:2002ds} K.~Abe {\it et al.} [Belle Collaboration],
Phys.\ Rev.\ Lett.\  {\bf 88}, 181803 (2002).

\bibitem{Wang:2007as} 
M.Z.~Wang {\it et al.} [Belle Collaboration], Phys.\ Rev.\ D {\bf 76}, 052004 (2007).

\bibitem{Aaij:2017vnw} 
R.~Aaij {\it et al.} [LHCb Collaboration],
Phys.\ Rev.\ Lett.\  {\bf 119}, 041802 (2017).

\bibitem{Wei:2007fg} 
J.T.~Wei {\it et al.} [Belle Collaboration], Phys.\ Lett.\ B {\bf 659}, 80 (2008).

\bibitem{Bevan:2014iga} 
A.J.~Bevan {\it et al.} [BaBar and Belle Collaborations],
Eur.\ Phys.\ J.\ C {\bf 74}, 3026 (2014). 

\bibitem{2b_baryonic}
Please consult ``17.12 B decays to baryons''  in Ref.~\cite{Bevan:2014iga}.

\bibitem{Suzuki:2006nn} 
M.~Suzuki,
J.\ Phys.\ G {\bf 34}, 283 (2007). 

\bibitem{Aaij:2013fla} 
R.~Aaij {\it et al.} [LHCb Collaboration],
Phys.\ Rev.\ D {\bf 88}, 052015 (2013). 

\bibitem{Aaij:2014tua} 
R.~Aaij {\it et al.} [LHCb Collaboration],
Phys.\ Rev.\ Lett.\  {\bf 113}, 141801 (2014). 

\bibitem{Beneke} 
M.~Beneke, G.~Buchalla, M.~Neubert and C.~T.~Sachrajda,
Nucl.\ Phys.\ B {\bf 591}, 313 (2000); 
245 (2001). 

\bibitem{Cheng:2001tr} 
H.Y.~Cheng and K.C.~Yang,
Phys.\ Rev.\ D {\bf 66}, 014020; 
094009 (2002). 

\bibitem{ali} A. Ali, G. Kramer, and C.D. Lu, Phys. Rev.  ${\bf D 58}$, 094009 (1998).


\bibitem{Buras:1998raa} 
A.J.~Buras, hep-ph/9806471.



\bibitem{Beneke:2002jn}
M.~Beneke and M.~Neubert,
Nucl.\ Phys.\ B {\bf 651}, 225 (2003).

\bibitem{BSW}
M. Wirbel, B. Stech, and M. Bauer, Z. Phys. C{\bf 29}, 637 (1985); {\bf 34}, 103 (1987);
M.~Bauer and M.~Wirbel, Z.\ Phys.\  C {\bf 42}, 671 (1989).

\bibitem{MFD} 
D.~Melikhov and B.~Stech, Phys.\ Rev.\ D {\bf 62}, 014006 (2000).


\bibitem{Brodsky:1973kr} 
S.J.~Brodsky and G.R.~Farrar, Phys.\ Rev.\ Lett.\  {\bf 31}, 1153 (1973);
Phys.\ Rev.\ D {\bf 11}, 1309 (1975).

\bibitem{Brodsky:2003gs} 
S.J.~Brodsky, C.E.~Carlson, J.R.~Hiller and D.S.~Hwang, Phys.\ Rev.\ D {\bf 69}, 054022 (2004).

\bibitem{Belitsky:2002kj} 
A.V.~Belitsky, X.D.~Ji and F.~Yuan, Phys.\ Rev.\ Lett.\  {\bf 91}, 092003 (2003).

\bibitem{Hsiao:2014zza} 
Y.K.~Hsiao and C.Q.~Geng, Phys.\ Rev.\ D {\bf 91}, 077501 (2015).

\bibitem{Aaij:2013fta} 
R.~Aaij {\it et al.} [LHCb Collaboration],
JHEP {\bf 10}, 005 (2013).

\bibitem{Aaij:2017gum} 
R.~Aaij {\it et al.} [LHCb Collaboration],
Phys.\ Rev.\ Lett.\  {\bf 119}, 232001 (2017). 

\bibitem{Fan:2012kn} 
Y.Y.~Fan, W.F.~Wang, S.~Cheng and Z.J.~Xiao,
Phys.\ Rev.\ D {\bf 87}, 094003 (2013). 


\bibitem{Krankl:2015fha} 
S.~Krankl, T.~Mannel and J.~Virto,
Nucl.\ Phys.\ B {\bf 899}, 247 (2015). 

\bibitem{Cheng:2016shb} 
H.Y.~Cheng, C.K.~Chua and Z.Q.~Zhang,
Phys.\ Rev.\ D {\bf 94}, 094015 (2016). 


\bibitem{R1} Y.K.~Hsiao and C.Q.~Geng, Phys.\ Lett.\ B727, 168 (2013).
\bibitem{R2} H.Y.~Cheng, C.Q.~Geng and Y.K.~Hsiao, Phys. Rev. D89, 034005 (2014).
\bibitem{R_BABAR} P.~del Amo Sanchez {\it et al.}  [BABAR Collaboration], 
Phys.\ Rev.\ D {\bf 85}, 092017 (2012).
\bibitem{Cheng:2008ss} 
H.Y.~Cheng, H.n.~Li and K.F.~Liu,
Phys.\ Rev.\ D {\bf 79}, 014024 (2009). 


\bibitem{Duplancic:2015zna} 
G.~Duplancic and B.~Melic,
JHEP {\bf 1511}, 138 (2015). 

\end{thebibliography}
\end{document}